
\input phyzzx
\hfuzz 5pt
\vfuzz 2pt
\Pubnum={OU-TAP-9 \cr
         KUNS 1268}
\date={September 1994}
\titlepage
\title{Gravitational Waves From a Particle Orbiting around
a Rotating Black Hole: Post-Newtonian Expansion}

\author{Masaru Shibata}

\address{Department of Earth and Space Science,
Osaka University, Toyonaka, Osaka 560, Japan}

\andauthor{Misao Sasaki, Hideyuki Tagoshi, Takahiro Tanaka}
\address{Department of Physics, Kyoto University,
Kyoto 606-01, Japan}

\abstract{Using the Teukolsky and Sasaki-Nakamura equations
for the gravitational perturbation of the Kerr spacetime,
we calculate the post-Newtonian expansion of the
energy and angular momentum luminosities of gravitational waves
from a test particle orbiting around a rotating
black hole up through ${\rm P^{5/2}N}$ order
beyond the quadrupole formula.
We apply a method recently developed by Sasaki to the case of a rotating
 black hole.
We take into account a small inclination of the orbital plane
to the lowest order of the Carter constant.
The result to ${\rm P^{3/2}N}$ order is in agreement with
a similar calculation by Poisson as well as with the
standard post-Newtonian calculation by Kidder et al.
Using our result, we calculate the integrated phase of gravitational
waves from a neutron star-neutron star binary and a black hole-neutron
 star binary during their inspiral stage.
We find that, in both cases, spin-dependent terms in
the P$^2$N and P$^{5/2}$N corrections
are important to construct effective template
waveforms which will be used for future laser-interferometric
gravitational wave detectors.
}
\def\Ci{{\rm Ci}}
\def\Si{{\rm Si}}
\REF\ligo{R. E. Vogt, in
{\it Proceeding of the 6th Marcel Grossmann Meeting on General
Relativity}, eds. H. Sato and T. Nakamura
 (World Scientific, Singapore, 1991), 244; \nextline
A. Abramovici et al., Science {\bf 256}, 325 (1992); \nextline
K. S. Thorne, in {\it Proceedings of the 8th Nishinomiya-Yukawa Memorial
Symposium: Relativistic Cosmology}, ed. M. Sasaki
 (Universal Academy Press, Tokyo, 1994), p67.}

\REF\Virgo{C.~Bradaschia et al., Nucl. Instrum. and Method {\bf A289},
518 (1990).}

\REF\Nar{R. Narayan, T. Piran and A. Shemi,~Ap.~J.~{\bf 379}, L17 (1991);
 \nextline E. S. Phinney, Ap. J. {\bf 380}, L17 (1991).}

\REF\Cutler{C. Culter et al., Phys. Rev. Lett. {\bf 70}, 2984 (1993).}

\REF\mar{D. Markovic, Phys. Rev. {\bf D48}, 4738 (1993).}

\REF\shibata{M. Shibata, K. Nakao and T. Nakamura, Phys. Rev. {\bf D},
submitted.}

\REF\will{C. M. Will, in {\it Proceedings of the 8th Nishinomiya-Yukawa
Memorial Symposium: Relativistic Cosmology}, ed. M. Sasaki
(Universal Academy Press, Tokyo, 1994), p83, and references therein.}

\REF\pois{E. Poisson, Phys. Rev. {\bf D47}, 1497 (1993).}

\REF\Curt{C. Cutler, L. S. Finn, E. Poisson and
G. J. Sussman, Phys. Rev. {\bf D47}, \nextline 1511 (1993).}

\REF\Kidder{L.~E.~Kidder,~C.~M.~Will~and~A.~G.~Wiseman,
Phys. Rev. {\bf D47}, \nextline R4183 (1993).}

\REF\shi{M. Shibata, Phys. Rev. {\bf D48}, 663 (1993).}

\REF\poisa{E. Poisson, Phys. Rev. {\bf D48}, 1860 (1993).}

\REF\shiba{M. Shibata, Prog. Theor. Phys. {\bf 90}, 595 (1993).}

\REF\tagonak{H. Tagoshi and T. Nakamura, Phys. Rev. {\bf 49} 4016 (1994).}

\REF\tagosas{H. Tagoshi and M. Sasaki, Prog. Theor. Phys. ??? (1994)
to be published.}

\REF\blan{For example, T. Futamase, Phys. Rev. {\bf D28}, 2373 (1983);
L. Blanchet and T. Damour, Phil. Trans. R. Soc. London
{\bf A320}, 379 (1986).}

\REF\Teu{S.~A.~Teukolsky,~Ap.~J.~{\bf 185}, 635 (1973).}

\REF\Sas{
M. Sasaki and T.~Nakamura, Prog. Theor. Phys. {\bf 67}, 1788 (1982);
\nextline M. Sasaki and T.~Nakamura, Phys. Lett. {\bf 89A}, 68 (1982).
}

\REF\New{
E. T. Newman and R.~Penrose, J. Math. Phys. {\bf 7}, 863 (1966).}

\REF\press{W. H. Press and S. A. Teukolsky,
Ap. J. {\bf 185}, 649 (1973).}

\REF\sasaki{M. Sasaki, Prog. Theor. Phys. {\bf 92}, 17 (1994).}

\REF\Cart{B.~Carter, Phy.~Rev.~{\bf 174}, 1559 (1968).}

\REF\Poi{T. Apostolatos, D. Kennefick, A. Ori
and E. Poisson, Phys. Rev. {\bf D47}, \nextline 5376 (1993).}

\REF\memb{K. S. Thorne, R. M. Price and D. MacDonald, {\it
Black Holes: The Membrance Paradigm} (Yale University Press,
New Haven, 1986).}

\REF\Tay{J. H. Taylor and J. M. Weisberg,
Ap. J. {\bf 345}, 434 (1989).}

\REF\Ander{S. B. Anderson et al., Nature {\bf 346}, 42 (1990).}

\REF\Wol{A.~Wolszczan,~Nature~{\bf 350}, 688 (1991).}

\REF\Puls{J. H. Taylor, R. N. Manchester and G. Lyne,
Ap. J. Suppl. {\bf 88}, 529 (1993).}

\REF\bild{L.~Bildsten and C.~Cutler, Ap. J. {\bf 400}, 175 (1992).}


\chapter{Introduction}

The last stage of an inspiraling compact binary such as binary neutron
stars is one of the promising sources of gravitational waves for
the near-future laser-interferometric detectors such as LIGO\refmark\ligo
and VIRGO\rlap.\refmark\Virgo\
 The main reasons why it is most promising
are that the event rate is expected to be $\sim3$ events/yr within 200
Mpc from a statistical study\refmark\Nar
and that it is possible to theoretically predict the amplitude of
gravitational waves with good accuracy.
Such a binary is, moreover, not only
a strong source of gravitational waves,
but also has a possibility to become
a treasury of physics of neutron stars\rlap,\refmark\Cutler
cosmology\rlap,\refmark\mar theories of gravity\rlap,\refmark\shibata
 etc., provided we obtain the data about binaries such as
masses, spins, distance to the earth, and so on.
However, to obtain those data with sufficient accuracy,
it is necessary to construct theoretical
template waveforms whose phasing has a fractional accuracy
of less than $10^{-4}$\rlap.\refmark\Cutler\
Hence considerable efforts have been paid to construct such
theoretical templates\rlap.\refmark{\will-\tagosas}\

To calculate gravitational waves from an inspiraling compact binary, the
standard method employed is the post-Newtonian expansion (PNE) of
the Einstein equations\rlap,\refmark\will
in which the equations are expanded in terms of a small parameter
$v \sim (M/r)^{1/2}$, where
$M$ and $r$ are the total mass and orbital length-scale of the system,
respectively.
Despite much effort, however, calculations have been successful to
only a few orders in $v$ beyond the leading (Newtonian) order so far.
More fundamentally, the nature of PNE has not been clarified due to its
complexity; nobody knows the convergence property of PNE nor the
validity of the polynomial expansion in $v$\rlap.\refmark\blan\
Given this situation, it is highly desirable to have a method which is
complementary to the standard PNE. The perturbative study of a black
hole spacetime is one of such, in which
we consider gravitational waves radiated by a particle of mass
$\mu \ll M$ orbiting around a black hole. This method, though restricted
to the case of $\mu\ll M$, is very powerful because we can calculate
fully general relativistic corrections of
gravitational waves by means of relatively simple analyses. It is then
fairly straightforward to evaluate the post-Newtonian corrections which
are to be calculated in the standard PNE.
This direction of research was first done analytically by
Poisson\refmark\pois to P$^{3/2}$N order and numerically by
Cutler et al\rlap.\refmark\Curt\  to P$^{3}$N order.
Then a highly accurate numerical calculation was done by
Tagoshi and Nakamura\refmark\tagonak to P$^4$N order
 and an analytical calculation to the same order was done by
Tagoshi and Sasaki\refmark\tagosas which confirmed the result of
Tagoshi and Nakamura. In particular, the appearance of
$\log v$ terms in the energy flux at $O(v^6)$ (P$^3$N order)
and at $O(v^8)$ (P$^4$N order) is confirmed and it is clarified that
the accuracy of the energy flux to at least P$^3$N oder is necessary to
construct template waveforms for optimal use of the
 interferometric data.

In this paper, we extend the analysis of Tagoshi and
Sasaki\refmark{\tagosas} to
the case of circular orbits around a rotating black hole to see
the effect of spin.
Some analytical\refmark{\Kidder,\poisa} and
numerical analyses\refmark{\shi,\shiba} of its effect
at its leading order (i.e., to P$^{3/2}$N order) have been performed.
They found that it gives rise to a large error in the integrated phase of
 gravitational waves from inspiraling binaries for a typical value of the
spin angular momentum if it is ignored in the template.
The large effect of the ${\rm P^{3/2}N}$ order spin-orbit term
causes us great anxiety about the effects of higher order terms
due to the spin angular momentum since
the convergence property of the PNE has been found to be
slow\rlap.\refmark{\Cutler,\tagonak,\tagosas}\
Hence we study the effect of spin
to P$^{5/2}$N order in this paper.

The paper is organized as follows.
In section 2, we show the basic formulas to perform the PNE
in our perturbative approach.
First we show the PNE of the Teukolsky radial function\refmark\Teu using
the Sasaki-Nakamura equation\rlap.\refmark\Sas\
 We must also perform the PNE of
the angular equation, which is given in Appendix D.
Then we describe the PNE of the source terms.
We consider circular orbits around a Kerr black hole, that is,
those at constant Boyer-Lindquist radial coordinate.
These circular orbits are, however, not necessarily on the equatorial
plane and the emission rates of gravitational waves are different for
different orbital inclination angles (or more precisely,
different values of the Carter constant)\rlap.\refmark\shiba\
To see the leading effect of the orbital inclination,
we consider orbits with small inclination angles.
In section 3, the energy and angular momentum luminosities to
$O(v^5)$ beyond Newtonian are derived. In section 4, using
the results given in section 3, we estimate errors in
the accumulated phase of gravitational waves caused by the non-vanishing
spin angular momentum.
Section 5 is devoted to the summary.
Throughout this paper we use the units of $c=G=1$ and
$GM_{\odot}/c^2=1.477$km.
\chapter{Formulation}

To calculate gravitational radiation from a particle orbiting
around a Kerr black hole, we start with
the Teukolsky equation\rlap.\refmark\Teu\  We focus on the
radiation going out to infinity described by
the fourth Newman-Penrose quantity, $\psi_4$\rlap,\refmark\New\
which may be expressed as
$$
\psi_4=(r-ia\cos\theta)^{-4}\int d\omega
 e^{-i\omega t}\sum_{\ell,m}
{e^{im\varphi}\over\sqrt{2\pi}}\,
{}_{-2}S^{a\omega}_{\ell m}(\theta)R_{\ell m\omega}(r),
\eqn\psifour
$$
where $_{-2}S_{\ell m}^{a\omega}$ is the spheroidal harmonic function of
spin weight $s=-2$, which are normalized as
$$
\int_0^{\pi} |_{-2}S_{\ell m}^{a\omega}|^2 \sin\theta d\theta=1,
\eqn\Snorm
$$
and $\lambda$ is the eigenvalue.
The radial function $R_{\ell m\omega}(r)$ obeys the Teukolsky equation
with spin weight $s=-2$,
$$
\Delta^2 {d \over dr}\Bigl({1 \over \Delta}{dR_{\ell m\omega} \over dr}
\Bigr)
-V(r)R_{\ell m\omega}=T_{\ell m\omega}(r),\eqno\eq
$$
where $T_{\ell m\omega}(r)$
is the source term whose explicit form will be shown later, and
$\Delta=r^2-2Mr+a^2$. The potential $V(r)$ is given by
$$
V(r)=-{K^2+4i(r-M)K \over \Delta}+8i\omega r+\lambda,\eqno\eq
$$
where $K=(r^2+a^2)\omega-ma$.

The solution of the Teukolsky equation at infinity
($r\rightarrow\infty$) is expressed as
$$
R_{\ell m\omega}(r)
 \rightarrow{r^3e^{i\omega r^*} \over 2i\omega B^{in}_{\ell m\omega}}
\int^{\infty}_{r_+}dr'{T_{\ell m\omega}(r') R_{\ell m\omega}^{in}(r')
\over\Delta^{2}(r')}
\equiv \tilde Z_{\ell m\omega}r^3e^{i\omega r^*},
\eqn\Rinfty
$$
where $r_+=M+\sqrt{M^2-a^2}$ denotes the radius of the event horizon
and $R_{\ell m\omega}^{in}$ is the homogeneous solution
which satisfies the ingoing-wave boundary condition at horizon,
$$
R_{\ell m\omega}^{in} \rightarrow\cases{
\displaystyle D_{\ell m\omega}\Delta^2  e^{-ikr^*}\,;&
$r^*\rightarrow -\infty,$\cr
\displaystyle r^3 B_{\ell m\omega}^{out}e^{i\omega r^*}+
r^{-1}B_{\ell m\omega}^{in} e^{-i\omega r^*}\,;&
$r^*\rightarrow +\infty,$ \cr
}\eqn\Rin
$$
where $k=\omega-ma/2Mr_+$ and
$r^*$ is the tortoise coordinate defined by
$$
{dr^* \over dr}={r^2+a^2 \over \Delta}.\eqno\eq
$$
For definiteness, we fix the integration constant such that
$r^*$ is given explicitly by
$$
\eqalign{
r^{*}&=\int {dr^*\over dr}dr \cr
&=r+{2Mr_{+}\over {r_{+}-r_{-}}}\ln{{r-r_{+}}\over 2M}
-{2Mr_{-}\over {r_{+}-r_{-}}}\ln{{r-r_{-}}\over 2M}\cr},
\eqn\rst
$$
where $r_{\pm}=M\pm \sqrt{M^2-a^2}$.

Thus in order to calculate gravitational waves emitted to infinity from
a particle in circular orbits,
we need to know the explicit form of the source term $T_{\ell m\omega}(r)$,
which has support only at $r=r_0$ where $r_0$ is the orbital radius
in the Boyer-Lindquist coordinate,
the ingoing-wave Teukolsky function $R^{in}_{\ell m\omega}(r)$ at $r=r_0$,
and its incident amplitude $B^{in}_{\ell m\omega}$ at infinity.
We consider the expansion of these quantities in terms of a small parameter,
$v^2 \equiv M/r_0$. Note that $v$ is approximately
equal to the orbital velocity, but not strictly equal to it
in the case of $a \not= 0$.

In addition to these, we need to expand the
spheroidal harmonics and their eigenvalues in powers of $a\omega$.
Since $\omega=O(\Omega)$ where $\Omega$ is the orbital angular
velocity of the particle, we have $a\omega=O(M\omega)=O(v^3)$.
Thus the expansion in powers of $a\omega$ is also a part of PNE.
Note also that the spin parameter of the black hole $a$ does not have to
be small but can be of order $M$.
We defer this expansion of the spheroidal harmonics to Appendix D.

In the following,
we consider the PNE of the ingoing-wave Teukolsky function and
the source term separately. Since the angular eigenvalues $\lambda$
comes into play in the radial equation, we set
$$
 \lambda=\lambda_0+ a\omega \lambda_1+a^2\omega^2 \lambda_2+O(v^9),
\eqn\lamPNE
$$
where $\lambda_0=(\ell-1)(\ell+2)$ and
$\lambda_1=-2m(\ell^2+\ell+4)/(\ell^2+\ell)$ have been
obtained previously\rlap,\refmark\press\ and $\lambda_2$ is derived in
Appendix D.

\section{Homogeneous solution}

The PNE of the Teukolsky equation for a rotating black hole
was first performed by Poisson up to $O(v^3)$ beyond the
quadrupole formula\rlap.\refmark\poisa\
In his method, he introduced a parameter
$\epsilon \equiv 2M\omega \sim O(v^3)$ and expanded the equation in
terms of it to the first order in $\epsilon$,
since only the homogeneous solution to $O(\epsilon)$ is needed to
obtain the energy flux up to $O(v^3)$ level.
The purpose here is to obtain the higher terms,
$O(v^n)$ ($n>3$). Specifically in this paper we consider the
gravitational wave luminosity to $O(v^5)$. Hence we
need the homogeneous solution accurate up to $O(\epsilon^2)$.
However, it is very difficult to treat the Teukolsky equation itself
to obtain the higher order corrections because
the 0th order solutions (\ie, the kernel functions)
already have a quite complicated form\rlap.\refmark\poisa\
 Hence instead of the Teukolsky equation,
we use the Sasaki-Nakamura (SN) equation\rlap,\refmark{\Sas}\
which is obtained by a certain transformation of the Teukolsky equation.
The reason to use it is that the
SN equation is a generalization of the
Regge-Wheeler (RW) equation for $a=0$ to the case of
$a \not=0$ and hence has a more tractable structure. In addition,
we can make use of algebraic formulas for the PNE
of the RW equation developed by Poisson\refmark\pois and by
Sasaki\rlap.\refmark\sasaki\

The SN equation has the form,
$$
\Bigl[{d^2 \over dr^{*2}}-F(r){d \over dr^{*}}-U(r)\Bigr]
X_{\ell m\omega}=0,
\eqn\sneq
$$
where the explicit forms of $F$ and $U$ are given in Appendix A.
It is obtained by the transformation from $R_{\ell m\omega}$ to
$X_{\ell m\omega}$ as
$$
R_{\ell m\omega}={1 \over \eta}
\Bigl\{\Bigl(\alpha+{\beta_{,r} \over \Delta}\Bigr)\chi_{\ell m\omega}
-{\beta \over \Delta}\chi_{\ell m\omega,r} \Bigr\},\eqn\xtor
$$
where $\chi_{\ell m\omega}=X_{\ell m\omega} \Delta/(r^2+a^2)^{1/2}$, and
the functions $\alpha$, $\beta$ and $\eta$ are shown in Appendix A.
Conversely, we can express $X_{\ell m\omega}$ in terms of
$R_{\ell m\omega}$ as
$$
X_{\ell m\omega}=(r^2+a^2)^{1/2}\,r^2\,J_{-}J_{-}
\left[{1\over r^2}R_{\ell m\omega}\right], \eqn\rtox
$$
where $J_{-}=(d/)-i(K/\Delta)$.
Then the ingoing-wave solution $X_{\ell m\omega}^{in}$ which corresponds
to $R_{\ell m\omega}^{in}$ has the following asymptotic behavior,
$$
X_{\ell m\omega}^{in} \rightarrow\cases{
\displaystyle A_{\ell m\omega}^{out}e^{i\omega r^*}+
A_{\ell m\omega}^{in} e^{-i\omega r^*}\,;& $r^* \rightarrow \infty,$ \cr
\displaystyle
 C_{\ell m\omega} e^{-ik r^*}\,;& $r^* \rightarrow -\infty,$ \cr}
\eqn\boun
$$
where $A_{\ell m\omega}^{in}$, $A_{\ell m\omega}^{out}$ and
$C_{\ell m\omega}$ are respectively related to
$B_{\ell m\omega}^{in}$, $B_{\ell m\omega}^{out}$ and
$D_{\ell m\omega}$ defined in Eq.$\Rin$ as
$$
\eqalign{
B_{\ell m\omega}^{in}&=-{1\over 4\omega^2}A_{\ell m\omega}^{in}, \cr
B_{\ell m\omega}^{out}&=-{4\omega^2\over c_0}A_{\ell m\omega}^{out}, \cr
D_{\ell m\omega}&={1\over d_{\ell m\omega}}C_{\ell m\omega}, \cr}
\eqn\ainbin
$$
where $c_0$ is given in Eq.(A.3) of Appendix A and
$$
\eqalign{
d_{\ell m\omega}&=\sqrt{2Mr_+}[(8-24iM\omega-16M^2\omega^2)r_{+}^2 \cr
&+(12iam-16M+16amM\omega+24iM^2\omega)r_{+}
-4a^2m^2-12iamM+8M^2]. \cr}
\eqn\plm
$$

Now let us consider the PNE of $X^{in}_{\ell m\omega}$.
It is characterized by the ingoing-wave boundary condition at horizon,
$X^{in}_{\ell m\omega}\sim e^{-ikr^*}$ as $r^*\rightarrow -\infty$.
However, since $r^*$ cannot be expanded in terms of
$\epsilon$ at $r^*\rightarrow -\infty$, a naive expansion of the SN
equation in terms of $\epsilon$ would obscure the boundary condition
at horizon.
One prescription to circumvent this difficulty
has been suggested by Sasaki\refmark{\sasaki}
in the case of the Schwarzschild black hole, namely to separate out
the factor $e^{-\omega (r^*-r)}$ from $X^{in}_{\ell m\omega}$
from the beginning.
Here we generalize this method to the case of the Kerr black hole.

First we introduce the variable $z=\omega r$ and
$$
\eqalign{
z^*&=z+\epsilon\left[{z_{+}\over {z_{+}-z_{-}}}\ln(z-z_+)
-{z_{-}\over {z_{+}-z_{-}}}\ln(z-z_-)\right]\cr
&=\omega r^*+\epsilon\ln\epsilon\,, \cr}
\eqn\zst
$$
where $z_\pm=\omega r_\pm$. For later convenience, we also
introduce a non-dimensional parameter $q=a/M$. Hence, for example,
$a\omega={1\over 2}q\epsilon$.
$z_{\pm}={1\over2}\epsilon(1\pm \sqrt{1-q^2})$.
As mentioned before, $q$ is not necessarily very small but can be
of order unity.

Next we define a function $\phi(z)$ as
$$
\eqalign{
\phi(z) &=\int dr\left({K\over \Delta}-\omega\right)\cr
&=z^*-z-{\epsilon\over 2}mq{1\over {z_{+}-z_{-}}}
\ln{{z-z_{+}}\over {z-z_{-}}}\,,\cr}
\eqn\phase
$$
which generalizes the phase function $\omega(r^*-r)$ of
 the Schwarzschild case.
Note that $e^{-i\phi(z)}\sim e^{-ikr^*}$ as $r^*\rightarrow -\infty$ and
$\sim e^{-i\omega(r^*-r)}$ as $r^*\rightarrow +\infty$.
Then we set
$$
X^{in}_{\ell m\omega}=\sqrt{z^2+a^2\omega^2}\xi_{\ell m}(z)
\exp\left(-i\phi(z)\right).
\eqn\xixi
$$
By this prescription, it is easily to implement the ingoing-wave
boundary condition on $X^{in}_{\ell m\omega}$.

Inserting Eq.$\xixi$ into Eq.$\sneq$ and expanding it in terms of
$\epsilon=2M\omega$, we obtain
$$
L^{(0)}[\xi_{\ell m}]=\epsilon L^{(1)}[\xi_{\ell m}]
+\epsilon Q^{(1)}[\xi_{\ell m}]+\epsilon^2 Q^{(2)}[\xi_{\ell m}]
+O(\epsilon^3), \eqn\eps
$$
where $L^{(0)}$, $L^{(1)}$, $Q^{(1)}$ and $Q^{(2)}$ are differential
operators given by
$$
\eqalignno{
L^{(0)}&={d^2\over dz^2}+{2\over z}{d\over dz}+\left(1-{\ell(\ell+1)
\over z^2}\right),
&\eqnalign{\Lzero}\cr
 L^{(1)}&={1\over z}{d^2\over dz^2}
+\left( {{1}\over z^2}
+{2i\over z}\right){d\over dz}
-\left( {{4}\over {z^3}}-{i\over z^2}
+{1\over z}\right),
&\eq \cr
 Q^{(1)}&={{iq\lambda_1}\over 2z^2}{d\over dz}
-{{4imq}\over {l(l+1)z^3}}
+{4mq\over {l(l+1)z^2}}+{{\lambda_1q+2mq}\over 2z^2},
&\eq \cr}
$$
for arbitrary $\ell$ and
$$
\eqalign{
Q^{(2)}=&-{q^2\over 4z^2}{d^2\over dz^2}
-\left({{9 q^2-2m^2q^2-6imq}\over {36z^3}}
+{{i(2m^2q^2-9q^2)-18mq}\over 108z^2}
\right){d\over dz} \cr
&+\left({9q^2+5m^2q^2-3imq\over 9z^4}
+{{i(2m^2q^2-9q^2)}\over 54z^3} \right.\cr
&\left.~~~~~~~~~~~~~~~~~~~~~~-{{-63imq+81q^2-2m^2q^2}\over378z^2}
\right), \cr}  \eqn\second
$$
for $\ell=2$. We shall see that we do not need $Q^{(2)}$ for $\ell\geq3$
for the post-Newtonian order we consider in this paper (\ie, up to
$O(v^5)$ beyond Newtonian).
Expanding $\xi_{\ell m}$ in terms of $\epsilon$ as
$$
\xi_{\ell m}=\sum_{n=0}^{\infty}\epsilon^n \xi^{(n)}_{\ell m}(z),
\eqn\expa
$$
we obtain from Eq.$\eps$ the iterative equations,
$$
\eqalignno{
&L^{(0)}[\xi^{(0)}_{\ell m}]=0, &\eqnalign\zeroji\cr
&L^{(0)}[\xi^{(1)}_{\ell m}]=L^{(1)}[\xi^{(0)}_{\ell m}]
+Q^{(1)}[\xi^{(0)}_{\ell m}]\equiv W^{(1)}_{\ell m},& \eqnalign{\first}\cr
&L^{(0)}[\xi^{(2)}_{\ell m}]=L^{(1)}[\xi^{(1)}_{\ell m}]
+Q^{(1)}[\xi^{(1)}_{\ell m}]+Q^{(2)}[\xi^{(1)}_{\ell m}]
\equiv W^{(2)}_{\ell m}. &\eqnalign\second
\cr}
$$
The general solution to Eq.$\zeroji$ is immediately obtained as
$$
\xi^{(0)}_{\ell m}=\alpha_\ell^{(0)} j_l+\beta_\ell^{(0)} n_l.
\eqn\xizero
$$

Now we consider the boundary condition. The argument is parallel to
that given in Ref.[\sasaki] for the Schwarzschild case.
The condition that $X^{in}_{\ell m\omega}\sim e^{-ikr^*}$ as
$r^*\rightarrow-\infty$ implies
$\sqrt{z^2+\epsilon^2q^2/4}\, \xi_{\ell m}(z)$
should be regular at $z=\omega r_+\sim \epsilon$.
Since $\epsilon$ can be made arbitrarily small,
$\sqrt{z^2+\epsilon^2q^2/4}\,\xi^{(n)}_{\ell m}(z)$
should be no more singular than $O(z^{-n})$ at $z=0$.
Thus in particular we have $\xi^{(0)}_{\ell m}=\alpha^{(0)}_\ell j_l$.
For convenience, we set $\alpha^{(0)}_\ell=1$.
Taking into account of the behavior of the lowest order solution,
we then infer that $\sqrt{z^2+\epsilon^2q^2/4}\xi^{(n)}_{\ell m}(z)$
must be no more singular than $z^{\ell+1-n}$ at $z=0$. Hence
for $n\leq 2$, the boundary condition is that
$\xi^{(n)}_{\ell m}$ is regular at $z=0$.

As noted previously, in the case of a circular orbit of radius $r_0$,
the source term $T_{\ell m\omega}$ has support only at $r=r_0$ and
$\omega r_0=O(\Omega r_0)=O(v)$.
Hence we only need $X^{in}_{\ell m\omega}$ at
$z=O(v)\ll1$ to evaluate the source integral, apart from the value
of the incident amplitude $A^{in}_{\ell m\omega}$.
Hence the PNE of $X^{in}_{\ell m\omega}$ corresponds to the expansion
not only in terms of $\epsilon=O(v^3)$ but also
$z$ by assuming $\epsilon\ll z\ll 1$.
In order to evaluate the gravitational
wave luminosity to $O(v^5)$ beyond the leading order,
we must calculate the series expansion of $\xi^{(n)}_{\ell m}$ in powers
of $z$ for $n=0$ to $\ell=4$, for $n=1$ to $\ell=3$
and for $n=2$ to $\ell=2$ (see Appendix C).
On the other hand, the accuracy of $A^{in}_{\ell m\omega}$ we need for
this purpose is $O(\epsilon)$ or $n\leq1$.
Thus we need the asymptotic behavior of $\xi^{(n)}_{\ell m}$ at
infinity only for $n=1$ (that for $n=0$ is trivially obtained).

To calculate $\xi_{\ell m\omega}$ to the accuracy discussed above,
we rewrite Eqs.\first\ and \second\ in the indefinite integral form
by using the spherical Bessel functions $j_\ell$ and $n_\ell$,
$$
\xi^{(n)}_{\ell m}=n_\ell\int^z dz z^2 j_\ell W^{(n)}_{\ell m}-
j_\ell\int^z dz z^2 n_\ell W^{(n)}_{\ell m}\quad(n=1,2). \eqn\green
$$
The series expansion formulas for $\xi^{(n)}_{\ell m}$ around $z=0$ is
easily obtained if we know those for $W^{(n)}_{\ell m}$.
Then it is straightforward to impose the boundary condition at $z=0$.
Further, if the integrations can be done exactly in closed analytic form,
it is easy to extract out the values of the incident amplitude
$A^{in}_{\ell m\omega}$ by examining the asymptotic
behavior at infinity of thus obtained functions. As we shall immediately
see, this is easily done for $n=1$, which is sufficient for our purpose.

For $n=1$, if we set $q=0$, Eq.$\first$ becomes the
same equation as that discussed in Ref.[\sasaki].
Hence the only correction is to include the contribution from
the $Q^{(1)}[\xi^{(0)}_{\ell m}]$ term.
Using the formulas developed in that paper,
this is easily done to yield
$$
\eqalign{
\xi^{(1)}_{\ell m} &=\alpha^{(1)}_{\ell }j_\ell
+{(\ell -1)(\ell +3)\over 2(\ell +1)(2\ell +1)}
j_{\ell +1}-{{\ell ^2-4}\over 2\ell (2\ell+1)}j_{\ell-1} \cr
&+z^2(n_\ell j_0-j_\ell n_0)j_0
+\sum_{k=1}^{\ell-1}\left({1\over k}+{1\over k+1}\right)
z^2(n_\ell j_k-j_\ell n_k)j_k \cr
&+n_\ell({\Ci} 2z-\gamma-\ln 2z )-j_\ell\Si 2z +ij_\ell\ln z \cr
&+{imq\over 2} \left({\ell^2+4}\over \ell^2(2\ell+1)\right)j_{\ell-1}
+{imq\over 2} \left({(\ell+1)^2+4}\over {(\ell+1)^2(2\ell+1)}\right)
j_{\ell+1},\cr} \eqn\xsi
$$
where ${\Ci(x)}=-\int^{\infty}_x dt\cos t/t$ and
${\Si(x)}=\int^x_0 dt\sin t/t$ are cosine and sine integral functions,
$\gamma$ is the Euler constant, and
$\alpha^{(1)}_\ell$ is an integration constant which represents the
arbitrariness of the normalization of
$X_{\ell m\omega}^{in}$. We set $\alpha^{(1)}_\ell=0$ for simplicity.

In the actual calculation of the gravitational radiation to infinity,
 we need to know $X^{in}_{\ell m\omega}(z)$ at $\epsilon\ll z\ll 1$.
 Using Eq.$\phase$, we obtain the expansion of Eq.$\xixi$ as
$$
\eqalign{X^{in}_{\ell m\omega}
&=\sqrt{z^2+a^2\omega^2}\,\xi_{\ell m}
\exp\left(-i\varphi \right) \cr
&=z\xi^{(0)}_{\ell m}+\epsilon\left(z\xi^{(1)}_{\ell m}
-{imq\over2}\xi^{(0)}_{\ell m}-iz\xi^{(0)}_{\ell m}\ln z\right)\cr
&+\epsilon^2\left[z\left(\xi^{(2)}_{\ell m}-i\xi^{(1)}_{\ell m}\ln z
-{1\over 2}\xi^{(0)}_{\ell m}(\ln z)^2\right) \right.\cr
&\left.+i\xi^{(0)}_{\ell m}-{i\over 2} \xi^{(1)}_{\ell m}mq
+{1\over z}\xi^{(0)}_{\ell m}\left(-{i\over 4}mq+{q^2\over 8}
-{{m^2q^2}\over 8}\right)-{1\over 2}\xi^{(0)}_{\ell m}mq\ln z\right].
\cr}\eqn\xin
$$

For $n=2$, expanding Eq.$\xsi$ in terms of $z$ and inserting it into
Eq.$\green$, we have the series expansion of $\xi^{(2)}_{\ell m}$.
Inserting it into Eq.$\xin$, we obtain
$$
\eqalignno{
X^{in}_{2m\omega}&={z^3 \over 15}-{z^5 \over 210}+{z^7 \over 7560}+O(z^9)
+\epsilon \Bigl({imqz^2 \over 30}-{13 z^4 \over 630}
-{11imqz^4 \over 3780}+O(z^6)\Bigr)
&\cr
&+\epsilon^2 \Bigl({q^2+2imq-m^2q^2 \over 120}z
-{{mq} \over 30}z^2+O(z^3)\Bigr),
&\eq\cr
X^{in}_{3m\omega}&={z^4 \over 105}-{z^6 \over 1890}+O(z^8)
+\epsilon
\Bigl(-{z^3 \over 126}+{2imq \over 945}z^3+O(z^5)\Bigr),
&\eq\cr
X^{in}_{4m\omega}&={z^5 \over 945}+O(z^7).&\eq\cr}
$$
Once we have the homogeneous solutions of the SN equation,
we have only to perform the transformation $\xtor$ to obtain
the corresponding solutions of the Teukolsky equation. The results are
$$\eqalignno{
\omega R^{in}_{2m\omega}&=
{{{z^4}}\over {30}} + {i\over {45}}\,{z^5}
- {{11\,{z^6}}\over {1260}} -   {i\over {420}}\,{z^7}
+ {{23\,{z^8}}\over {45360}} + {i\over {11340}}\,{z^9}&\eqnalign\elltwo\cr
&+\epsilon\Bigl(
{{-{z^3}}\over {15}} - {i\over {60}}\,m\,q\,{z^3}
- {i\over {60}}\,{z^4} + {{m\,q\,{z^4}}\over {45}}& \cr
&- {{41\,{z^5}}\over {3780}} +   {{277\,i}\over {22680}}\,m\,q\,{z^5}
- {{31\,i}\over {3780}}\,{z^6} -  {{7\,m\,q\,{z^6}}\over {1620}}
\Bigr)&\cr
&+\epsilon^2\Bigl(
{{{z^2}}\over {30}} + {i\over {40}}\,m\,q\,{z^2} +
  {{{q^2}\,{z^2}}\over {60}} - {{{m^2}\,{q^2}\,{z^2}}\over {240}} &\cr
&- {i\over {60}}\,{z^3} - {{m\,q\,{z^3}}\over {30}}
+ {{i}\over {90}}\,{q^2}\,{z^3}
- {i\over {120}}\,{m^2}\,{q^2}\,{z^3}\Bigr),
&\eq\cr
\omega R^{in}_{3m\omega}&=
{{{z^5}}\over {630}} + {i\over {1260}}\,{z^6} - {{{z^7}}\over {3780}}
- {i\over {16200}}\,{z^8} &\cr
&+\epsilon\Bigl(
{{-{z^4}}\over {252}} - {i\over {1890}}\,m\,q\,{z^4}
- {i\over {756}}\,{z^5} + {{11\,m\,q\,{z^5}}\over {22680}}
\Bigr),
&\eq\cr
\omega R^{in}_{4m\omega}&={z^6\over 11340}+{iz^7\over 28350}.
&\eq\cr}
$$
Here, it is worth noting that the terms linear in $q(=a/M)$ at $O(v^2)$
beyond the leading term in each $R^{in}_{\ell m\omega}$ are pure imaginary.
This implies
there is no linear term in $q$ at the P$^1$N order of
the luminosity.  Such terms will appear at the ${\rm P^{3/2}N}$ order,
which is in fact what was found by Poisson\rlap.\refmark\poisa\
Further, it is expected that terms linear in $q$ will not
appear at the P$^2$N order, but at the P$^{5/2}$N order.
By the same argument, terms quadratic in $q$ are expected to
appear at the ${\rm P^2N}$ order, but not at the
${\rm P^{5/2}N}$ order.

Next, we consider $A_{\ell m\omega}^{in}$ to $O(\epsilon)=O(v^3)$.
The procedure is the same as that in the Schwarzschild
case\rlap.\refmark\sasaki\  Using the relations
$j_{\ell+1}\sim-j_{\ell-1}\sim(-1)^{\ell+n}n_{2n-\ell}$,
etc., we obtain the asymptotic behavior of $\xi^{(1)}_{\ell m}$
at $z=\infty$ as
$$
\xi^{(1)}_{\ell m}\sim -{\pi\over 2}j_\ell
+\left(q^{(1)}_{\ell m}-\ln z\right)n_\ell+ij_\ell\ln z,
\eqn\xiasym
$$
where
$$
q^{(1)}_{\ell m}={1\over 2}\left[\psi(\ell)+\psi(\ell+1)
+{{(\ell-1)(\ell+3)}\over \ell(\ell+1)}\right]-\ln 2
-{{2imq}\over {\ell^2(\ell+1)^2}},
\eqn\qichi
$$
and $\psi(\ell)$ is the digamma function,
$$
\psi(\ell)=\sum^{\ell-1}_{k=1}{1\over k}-\gamma.
$$

To consider the asymptotic form of $X^{in}_{\ell m\omega}$,
we set $\xi^{(1)}_{\ell m}=f^{(1)}_{\ell m}+ij_{\ell}\ln z$ and
express the asymptotic form of
$f^{(1)}_{\ell m}$ at $z\rightarrow\infty$ as
$$
\eqalign{
f^{(1)}_{\ell m}&\rightarrow P^{(1)}_{\ell m}j_\ell+Q^{(1)}_{\ell m}n_\ell
\cr
&={1\over 2}\left(P^{(1)}_{\ell m}-iQ^{(1)}_{\ell m}\right)h^{(1)}_\ell
+{1\over 2}\left(P^{(1)}_{\ell m}+iQ^{(1)}_{\ell m}\right)h^{(2)}_\ell,\cr}
$$
where $h^{(1)}_\ell$ and $h^{(2)}_\ell$ are the spherical Hankel functions
of the first and second kinds, respectively, which are given by
$$
h^{(1)}_\ell(z)=j_\ell(z)+in_\ell(z)\rightarrow(-i)^{\ell+1}{e^{iz}\over z},
{}~h^{(2)}_\ell(z)=j_\ell(z)-in_\ell(z)\rightarrow i^{\ell+1}{e^{-iz}\over z}.
\eqno\eq
$$
\ From Eq.\xiasym, we have
$$
 P^{(1)}_{\ell m}=-{\pi\over2}\,,\qquad
 Q^{(1)}_{\ell m}=q^{(1)}_{\ell m}-\ln z\,.
\eqn\PandQ
$$
Then the asymptotic form of $X^{in}_{\ell m\omega}$ becomes
$$
\eqalign{
X^{in}_{\ell m\omega}
&=\sqrt{z^2+a^2\omega^2}\xi_{\ell m}\exp(-i\varphi) \cr
&\sim z\left[j_\ell+\epsilon(f^{(1)}_{\ell m}+ij_\ell\ln z)+\ldots\right]
\exp\Bigl[(-i\Bigl(z^*-z-{mq\over 2\sqrt{1-q^2}}
\ln{{z-z_+}\over {z-z_-}}\Bigr)\Bigr]\cr
&\sim{1\over 2}e^{-iz^*}(ze^{iz}h^{(2)}_{\ell m})
\left[1+\epsilon\bigl(P^{(1)}_{\ell m}+i(Q^{(1)}_{\ell m}+\ln z)\bigr)
+\ldots\right]\cr
&~~~~+{1\over 2}e^{iz^*}(ze^{-iz}h^{(1)}_{\ell m})
\left[1+\epsilon\bigl(P^{(1)}_{\ell m}-i(Q^{(1)}_{\ell m}+\ln z)\bigr)
+\ldots\right].\cr
} \eqn\xina
$$
\ From this equation, $A^{in}_{\ell m\omega}$ can be easily extracted out:
$$
A^{in}_{\ell m\omega}={1\over 2}i^{\ell+1}e^{-i\epsilon\ln\epsilon}
\left[1+\epsilon\left(-{\pi\over 2}+iq^{(1)}_{\ell m}\right)
+\ldots\right], \eqn\aina
$$
where we use the fact that $\omega r^*=z^*-\epsilon\ln\epsilon$
from our definition of $z^*$.
Specifically for $\ell=2$, 3 and 4, to the orders respectively
required, we have
$$
\eqalignno{
&A^{in}_{2m\omega}=
-{i \over 2}\Bigl\{ 1-\epsilon{\pi\over 2}
+i\epsilon \Bigl({5 \over 3}-\gamma -\log 2\Bigr)+
{mq \over 18}\epsilon \Bigr\}+O(\epsilon^2),
&\eq\cr
&A^{in}_{3m\omega}=
{1 \over 2}\Bigl\{ 1-\epsilon{\pi\over 2}
+i\epsilon \Bigl({13 \over 6}-\gamma -\log 2\Bigr)+
{mq \over 72}\epsilon \Bigr\}+O(\epsilon^2),
&\eq\cr
&A^{in}_{4m\omega}={i \over 2}+O(\epsilon).
&\eq\cr}
$$
Finally, the corresponding incident ampliutudes
$B^{in}_{\ell m\omega}$ for the Teukolsky function are
obtained from Eq.$\ainbin$.

\section{The source term}

A test particle obeys the equations of motion,
$$\eqalign{
&\Sigma{d\theta \over d\tau}=\pm\Bigl[C-\cos^2\theta \Bigl\{a^2(1-E^2)+
{l_z^2 \over \sin^2\theta}\Bigr\}\Bigr]^{1/2} \equiv \Theta(\theta),\cr
&\Sigma{d\varphi \over d\tau}=-\Bigl(aE-{l_z \over \sin^2\theta}\Bigr)
+{a \over \Delta}\Bigl(E(r^2+a^2)-al_z\Bigr),\cr
&\Sigma{dt \over d\tau}=
-\Bigl(aE-{l_z \over \sin^2\theta}\Bigr)a\sin^2\theta
+{r^2+a^2 \over \Delta}\Bigl(E(r^2+a^2)-al_z\Bigr),\cr
&\Sigma{dr \over d\tau}=\pm\sqrt{R},\cr}\eqn\test
$$
where $E$, $l_z$ and $C$ are the energy, the $z$-component of the
angular momentum and the Carter constant\refmark\Cart of a test particle,
respectively\rlap,\footnote{*}{In this section, these constants of
motion are those measured in units of $\mu$.
That is, if expressed in the standard units, $E$, $l_z$ and $C$ in
Eq.\test\ are to be replaced with $E/\mu$, $l_z/\mu$ and $C/\mu^2$,
respectively.}\  $\Sigma=r^2+a^2\cos^2\theta$ and
$$
R=[E(r^2+a^2)-al_z]^2-\Delta[(Ea-l_z)^2+r^2+C].\eqno\eq
$$
We consider the case in which a particle moves along
a constant radius $r=r_0$, but precesses around the symmetric axis.
The degree of precession is determined by the value of $C$.
If $r_0$ and $C$ are given,
the energy $E$ and the z-component of the angular momentum $l_z$
are obtained by the two equations,
$R=0$ and $dR/dr=0$.

Then the energy momentum tensor of a test particle is written as
$$
T^{\mu\nu}={\mu \over \Sigma\sin\theta dt/d\tau}
{dz^{\mu} \over d\tau}{dz^{\nu} \over d\tau}
\delta(r-r_0)\delta(\theta-\theta(t))\delta(\varphi-\varphi(t)).
\eqno\eq
$$
The source term of the Teukolsky equation is
$$
T_{\ell m\omega}
=-4\int d\Omega dt\rho^{-5}{\overline\rho}^{-1}(B_2'+B_2'^*)
e^{-im\varphi+i\omega t}{_{-2}S^{a\omega}_{\ell m} \over \sqrt{2\pi}},
\eqn\teuq
$$
where
$$\eqalign{
B_2'=&-{1 \over 2}\rho^8{\overline \rho}L_{-1}[\rho^{-4}L_0
(\rho^{-2}{\overline \rho}^{-1}T_{nn})]\cr
&-{1 \over 2\sqrt{2}}\rho^8{\overline \rho}\Delta^2 L_{-1}[\rho^{-4}
{\overline \rho}^2 J_+(\rho^{-2}{\overline \rho}^{-2}\Delta^{-1}
T_{{\overline m}n})],\cr
B_2'^*=&-{1 \over 4}\rho^8{\overline \rho}\Delta^2 J_+[\rho^{-4}J_+
(\rho^{-2}{\overline \rho}T_{{\overline m}{\overline m}})]\cr
&-{1 \over 2\sqrt{2}}\rho^8{\overline \rho}\Delta^2 J_+[\rho^{-4}
{\overline \rho}^2 \Delta^{-1} L_{-1}(\rho^{-2}{\overline \rho}^{-2}
T_{{\overline m}n})],\cr}\eqn\teu
$$
with
$$
\eqalign{
\rho&=(r-ia\cos\theta)^{-1},\cr
L_s&=\partial_{\theta}+{m \over \sin\theta}
-a\omega\sin\theta+s\cot\theta,\cr
J_+&=\partial_r+{iK \over \Delta},\cr}
\eqno\eq
$$
and ${\overline Q}$ denoting the complex conjugate of $Q$.
In the present case, the tetrad components of the energy momentum
tensor, $T_{n\,n}$, $T_{{\overline m}\,n}$ and
$T_{{\overline m}\,{\overline m}}$, are in the form,
$$\eqalign{
&T_{n\,n}={C_{n\,n} \over \sin\theta}
\delta(r-r_0) \delta(\theta-\theta(t)) \delta(\varphi-\varphi(t)),\cr
&T_{{\overline m}\,n}={C_{{\overline m}\,n} \over \sin\theta}
\delta(r-r_0) \delta(\theta-\theta(t)) \delta(\varphi-\varphi(t)),\cr
&T_{{\overline m}\,{\overline m}}=
{C_{{\overline m}\,{\overline m}} \over \sin\theta}
\delta(r-r_0) \delta(\theta-\theta(t)) \delta(\varphi-\varphi(t)),\cr}
\eqn\teut
$$
where
$$\eqalign{
&C_{n\,n}={\mu \over 4\Sigma^3 \dot t}[E(r^2+a^2)-al_z]^2,\cr
&C_{{\overline m}\,n}=
-{\mu \rho \over 2\sqrt{2}\Sigma^2 \dot t}[E(r^2+a^2)-al_z]
[i\sin\theta\Bigl(aE-{l_z \over \sin^2\theta}\Bigr)
+\Theta(\theta)],\cr
&C_{{\overline m}\,{\overline m}}=
{\mu \rho^2 \over 2\Sigma \dot t }
[i\sin\theta \Bigl(aE-{l_z \over \sin^2\theta}\Bigr)
+\Theta(\theta)]^2,\cr}
\eqno\eq
$$
and $\dot t=dt/d\tau$.

Substituting Eq.\teu\ into Eq.\teuq\ and performing integration by part,
we obtain
$$\eqalign{
T_{\ell m\omega}=&-{4 \over \sqrt{2\pi}}\int^{\infty}_{-\infty}
dt e^{i\omega t-im\varphi}
\Bigl[-{1 \over 2}L_1^{\dag} \bigl\{ \rho^{-4}L_2^{\dag}(\rho^3 S) \bigr\}
C_{n\,n}\rho^{-2}{\overline \rho}^{-1}\delta(r-r_0) \cr
&+{\Delta^2 {\overline \rho}^2 \over \sqrt{2} \rho}
\bigl(L_2^{\dag} S+ia({\overline \rho}-\rho)\sin\theta S\bigr)
J_+ \bigl\{ C_{{\overline m}\,n}\rho^{-2}{\overline \rho}^{-2}\Delta^{-1}
\delta(r-r_0) \bigr\} \cr
&+{1 \over 2\sqrt{2} }
L_2^{\dag}\bigl\{ \rho^3 S({\overline \rho}^2 \rho^{-4})_{,r} \bigr\}
C_{{\overline m}\,n}\Delta \rho^{-2}{\overline \rho}^{-2}
\delta(r-r_0) \cr
&-{1 \over 4}\rho^3 \Delta^2 S J_+\bigl\{\rho^{-4}
J_+\bigl({\overline \rho} \rho^{-2}C_{{\overline m}\,{\overline m}}
\delta(r-r_0)\bigr) \bigr\}
\Bigr],\cr
\equiv &\int^{\infty}_{-\infty} dt\, e^{i\omega t-im\varphi}\,
t_{\ell m\omega},
\cr}\eqn\tass
$$
where
$$
L_s^{\dag}=\partial_{\theta}-{m \over \sin\theta}
+a\omega\sin\theta+s\cot\theta,\eqno\eq
$$
and $S$ denotes $_{-2}S_{\ell m}^{a\omega}(\theta (t))$ for simplicity.

Equation \tass\ can be further simplified by noting that
the orbits of our interest have the properties,
$$
\theta(t+\Delta t)=\theta(t),~~~~~~~~
\varphi(t+\Delta t)=\varphi(t)+\Delta \varphi,
\eqno\eq
$$
where $\Delta t$ is the orbital period (of the motion in the
$\theta$-direction)
and $\Delta \varphi$ is the phase advancement during $\Delta t$.
For convenience, we set
$$
 \Omega_\theta\equiv {2\pi\over\Delta t}\,,\quad
 \Omega_\varphi\equiv {\Delta\varphi\over\Delta t}\,.
\eqn\Omegadef
$$
Then the source term reduces to the form,
$$
T_{\ell m\omega}=
T_{one} \sum_k e^{ik(\omega \Delta t-m \Delta \varphi)}
=\Omega_{\theta}T_{one}\sum_n \delta(\omega-\omega_n),
\eqn\sour
$$
where
$$
\eqalignno{
\omega_n=
&n\Omega_\theta+m\Omega_{\varphi}\,,\quad
(n=0,\pm 1,\pm 2,\cdots),&\eq\cr
T_{one}=
&\int^{\Delta t}_0 dt e^{i\omega t-im\varphi (t)}t_{\ell m\omega}
&\cr
=&\Delta^{2}
\Bigl[(A_{n\,n\,0}+A_{{\overline m}\,n\,0}
+A_{{\overline m}\,{\overline m}\,0})
\delta(r-r_0)
&\cr
&+\left\{(A_{{\overline m}\,n\,1}+A_{{\overline m}\,{\overline m}\,1})
\delta(r-r_0)\right\}_{,r}
+\left\{A_{{\overline m}\,{\overline m}\,2}\delta(r-r_0)\right\}_{,rr}
\Bigr],
&\eqnalign\Tone\cr}
$$
and the $A$'s are given in Appendix B. Inserting Eq.\sour\ into
Eq.\Rinfty, we obtain $\tilde Z_{\ell m\omega}$ as
$$\eqalign{
\tilde Z_{\ell m\omega}=&\sum_{n}\delta(\omega-\omega_n)
      Z_{\ell m\omega_n}\,;\cr
 Z_{\ell m\omega_n}=
&{\Omega_\theta\over2i\omega_n B^{in}_{\ell m\omega_n}}
\Bigl[R^{in}_{\ell m\omega_n}\{A_{n\,n\,0}+A_{{\overline m}\,n\,0}
+A_{{\overline m}\,{\overline m}\,0}\}
\cr
&-{dR^{in}_{\ell m\omega_n} \over dr}\{ A_{{\overline m}\,n\,1}
+A_{{\overline m}\,{\overline m}\,1}\}
 +{d^2 R^{in}_{\ell m\omega_n} \over dr^2}
 A_{{\overline m}\,{\overline m}\,2}\Bigr]_{r=r_0}\,.
\cr}
\eqn\zzq
$$

Now, let us consider the orbital integral in Eq.\Tone.
To perform them, we must know the trajectory of a test particle, \ie,
$\theta(t)$ and $\varphi(t)$, but they are not simple
 analytical functions of $t$ for general trajectories.
Thus, we here consider the case in
which a particle is in the orbit with small inclination.
To be specific, we introduce a dimensionless parameter $y$ defined by
$$
y={C\over Q^2}\,;\quad Q^2=l_z^2+a^2(1-E^2),
\eqn\ydef
$$
and regard it as small.
Since $Q^2\sim l_z^2$ and
$C \sim l_x^2+l_y^2$\rlap,\refmark\shiba\
this is equivalent to assuming $l_x^2+l_y^2 \ll l_z^2$.
Note also that we do not need the exact expressions for $E$ and $l_z$
in terms of $r_0$ and $C$ (or $y$), but only the PNE of them.
To the first order of $y$ as well as to the P$^{5/2}$N order,
they are given by
$$
\eqalign{
E=&1-{M \over 2r_0}+{3M^2 \over 8r_0^2}
-{M^{3/2}a \over r_0^{5/2}}(1-{y\over2})+O(v^6),\crr
l_z=&(Mr_0)^{1/2}\Biggl[\Bigl(1-{y \over 2}\Bigr)
+{3 M\over 2r_0}(1-{y\over2})-{3M^{1/2}a \over r_0^{3/2}}(1-y)\cr
&\quad+{27M^2\over8 r_0^2}(1-{y\over2})
+{a^2\over r_0^2}(1-2y)
-{15 M^{3/2}a \over 2r_0^{5/2}}(1-y)+O(v^6)\Biggr].\cr}
\eqno\eq
$$

To solve the geodesic equations under the assumption $y\ll1$, we first
set $\theta=\pi/2+y^{1/2} \theta'$ and consider the geodesic equation for
$\theta$. It then becomes
$$
\Bigl({d\theta' \over d\tau}\Bigr)^2
={1\over\Sigma^2}\left[Q^2 - {\sin^2(y^{1/2} \theta')\over y}
\Bigl\{a^2(1-E^2)+{l_z^2 \over \cos^2(y^{1/2} \theta') }
\Bigr\}\right].\eqn\jkl
$$
Since the r.h.s of Eq.\jkl\ contains only even-functions of
$y^{1/2}\theta'$, we can solve it iteratively by
expanding $\theta'$ as
$$
\theta'=\theta_{(0)}+y\theta_{(1)}+y^2\theta_{(2)}+\cdots.\eqno\eq
$$
This is similar in spirit to the method used by
Apostolatos et al.\rlap,\refmark\Poi\
who considered gravitational waves from a particle in an elliptical
orbit around a non-rotating black hole with small eccentricity $e\ll1$.
However, here we only consider the lowest order solution $\theta_{(0)}$.
This means we take into account the effect of inclination up to $O(y)$,
as seen from the structure of the geodesic equations \test.
The equation for $\theta_{(0)}$ is
$$
\Bigl({d\theta_{(0)}\over d\tau}\Bigr)^2=
  {Q^2\over\Sigma^2}(1-\theta_{(0)}^2),
\eqno\eq
$$
or dividing it by $(dt/d\tau)^2$,
$$
\Bigl({d\theta_{(0)}\over dt}\Bigr)^2
={Q^2\over\sigma^2}(1-\theta_{(0)}^2),
\eqno\eq
$$
where
$$
\sigma\equiv -a(aE-l_z)+{a^2+r_0^2 \over \Delta(r_0)}
\bigl\{ E(r_0^2+a^2)-al_z \bigr\}.\eqno\eq
$$
Then the solution is easily obtained as
$$
\theta_{(0)}=\sin(\Omega_{\theta} t);\quad
\Omega_{\theta}={Q \over \sigma}\,\left(={2\pi \over \Delta t}\right),
\eqn\exi
$$
where we have chosen $\theta_{(0)}=0$ at $t=0$.
Thus we have
$$
\theta={\pi\over2}+y^{1/2}\sin(\Omega_{\theta}t).
\eqn\thesol
$$
Note that the solution \thesol\ implies that the inclination angle
$\theta_i$ is indeed given by $\theta_i=y^{1/2}$ in the present
approximation.

Next, we consider the geodesic equation for $\varphi$.
Taking account of the terms up to $O(y)$, it becomes
$$
\eqalign{
{d\varphi \over dt}
=&{\kappa \over \sigma}\left[1+\Bigl({l_z \over \kappa}
           -{a^2E\over \sigma}\Bigr)y\theta_{(0)}^2\right]
\cr
=&\Omega_\varphi-y{\Omega_2\over2}\cos(2\Omega_\theta t),
\cr}
\eqn\vap
$$
where
$$
\kappa \equiv -(aE-l_z)+{a \over \Delta(r_0)}
\{ E(r^2_0+a^2)-al_z \},
$$
and
$$
\Omega_{\varphi}={\kappa \over \sigma}+{1\over2}y\Omega_2\,
\left(={\Delta \varphi \over \Delta t}\right),
\quad
\Omega_2={\kappa \over \sigma}
\Bigl({l_z \over \kappa}-{a^2E \over \sigma}\Bigr).
\eqno\eq
$$
The solution to Eq.\vap\ with $\varphi=0$ at $t=0$ is
$$
\varphi=\Omega_{\varphi}t
 -y{\Omega_2 \over 4\Omega_{\theta}}\sin(2\Omega_{\theta} t).\eqn\solv
$$
Note that $\Omega_{\varphi} \not= \Omega_{\theta}$.
This means the precession of a test particle orbit around the
spin axis of the black hole.
Specifically, to the order required for the present purpose (see
\S 3 below), we have
$$\eqalign{
\Omega_{\varphi}&={M^{1/2} \over r_0^{3/2}}
\left[1-{M^{1/2}a \over r_0^{3/2}}
+{3 \over 2}y\Bigl({M^{1/2}a \over r_0^{3/2}}-{a^2 \over r_0^2}\Bigr)
 + O(v^6)\right],\cr
\Omega_{\theta}&={M^{1/2} \over r_0^{3/2}}
\left[1-{3M^{1/2}a \over r_0^{3/2}}+{3a^2 \over 2r_0^2}
+O(v^6)+O(y)\right].\cr}
\eqn\omeeq
$$
We see that $\Omega_{\varphi}-\Omega_{\theta} \rightarrow 2Ma/r_0^3$
for $r_0 \rightarrow \infty$ and $y \rightarrow 0$,
which is just the Lense-Thirring precessional
frequency\rlap.\refmark\memb\

Now that we have the solution of the geodesic equations,
we can estimate the $A$'s in Eq.\zzq.
Up to $O(y)$, they are integrals of the form (see
Appendix B),
$$\eqalign{
I_{\ell m\omega_n}=&\int^{\Delta t}_0 dt e^{i\omega_n t-im\varphi(t)}
\Bigl\{
Q_{\ell m\omega_n}^0+y^{1/2} Q_{\ell m\omega_n}^1\theta_{(0)}
+y Q_{\ell m\omega_n}^2\theta_{(0)}^2
\cr
&+y^{1/2}Q_{\ell m\omega_n}^3 {d\theta_{(0)}\over dt}
+yQ_{\ell m\omega_n}^4\theta_{(0)}{d\theta_{(0)}\over dt}
+yQ_{\ell m\omega_n}^5 \Bigl( {d\theta_{(0)}\over dt} \Bigr)^2 \Bigr\}
+O(y^{3/2}),\cr}\eqn\iieq
$$
where $Q_{\ell m\omega_n}^k$ for $k=1 \sim 5$ are complicated functions
of $r_0$.
Substituting Eqs.\exi, \solv\ into Eq.\iieq,  and using
the approximation,
$$
e^{i\omega_n t-im\varphi(t)} = e^{in\Omega_{\theta} t}
\Bigl(1+y{m \Omega_2 \over 8\Omega_{\theta}}
(e^{2i\Omega_{\theta} t}-e^{-2i\Omega_{\theta} t})+O(y^{3/2})\Bigr),
\eqno\eq
$$
we find
$$\eqalign{
I_{\ell m\omega_n}=
&{2\pi\over\Omega_\theta}\biggl[\Bigl\{ \delta_{n0}
+y{m\Omega_2\over 8\Omega_{\theta}}
(\delta_{n,-2}-\delta_{n,2}) \Bigr\}Q_{\ell m\omega_n}^0
+y^{1/2}{1\over 2i}(\delta_{n,-1}-\delta_{n,1})
 Q_{\ell m\omega_n}^1
\cr
&+y{1\over 4}(2\delta_{n,0}-\delta_{n,-2}-\delta_{n,2})
 Q_{\ell m\omega_n}^2
+y^{1/2}{\Omega_\theta\over 2}(\delta_{n,-1}+\delta_{n,1})
 Q_{\ell m\omega_n}^3
\cr
&+y{\Omega_\theta\over 4i}(\delta_{n,-2}-\delta_{n,2})Q_{\ell m\omega_n}^4
+y{\Omega_\theta^2\over 4}(2\delta_{n,0}+\delta_{n,-2}+\delta_{n,2})
 Q_{\ell m\omega_n}^5\biggr]
\cr
&+O(y^{3/2}),\cr}
\eqno\eq
$$
where $\delta_{n,n'}$ is Kronecker's delta.
Applying these formulas to $A$s of Eq.\zzq,
the amplitude $Z_{\ell m\omega_n}$ is found to have the form,
$$\eqalign{
Z_{\ell m\omega_n}=
&\Bigl[\bigl(Z^{0,0}+yZ^{0,2}) \delta_{n,0}
 +y^{1/2}\bigl(Z^{1,1}\delta_{n,1}+Z^{1,-1}\delta_{n,-1}\bigr)
\cr
&+y\bigl(Z^{2,2}\delta_{n,2}+Z^{2,-2}\delta_{n,-2}\bigr)+O(y^{3/2})\Bigr],
\cr}
\eqn\Zform
$$
where $Z^{i,j}$ are functions of $r_0$.
In principle, algebraic calculations of $Z^{i,j}$ are
straightforward, but they are almost impossible by hand in practice.
Thus to avoid trivial mistakes as well as to save time,
we have made use of the algebraic manipulation program,
{\it Mathematica}, to obtain $Z^{i,j}$.
We shall not list their explicit forms here since
they are too complicated and not much insight can be gained from them.

\chapter{The energy and angular momentum fluxes}

In this section, we calculate the energy and angular momentum fluxes
to $O(v^5)$ beyond the quadrupole formula and to $O(y)$ in the
 orbital inclination. From Eqs.\psifour,
$\psi_4$ at $r \rightarrow \infty$ to the required order takes the form,
$$
\psi_4={1\over r}\sum_{n=-2}^{2}\sum_{\ell=2}^{4}\sum_{m=-\ell}^{\ell}
Z_{\ell m\omega_n}{{}_{-2}S_{\ell m}^{a\omega_n} \over \sqrt{2\pi}}
e^{i\omega_n(r^*-t)+im\varphi}.\eqn\heq
$$
It is worth noting that there are symmetry relations,
${}_{-2}S_{\ell m}^{a\omega_n}(\theta)
= {}_{-2}S_{\ell-m}^{a\omega_{-n}}(\pi-\theta)$
and $Z_{\ell-m\omega_{-n}}
=(-1)^n{\overline Z}_{\ell m\omega_n}$.
At infinity, $\psi_4$ is related to the two independent modes of
gravitational waves $h_+$ and $h_\times$ as
$$
\psi_4={1\over2}(\ddot h_{+}-i\ddot h_{\times}).
\eqn\hplcr
$$

\ From Eqs.\heq\ and \hplcr, the energy flux averaged over
$t \gg \Delta t$ is given by
$$
\Bigl< {dE \over dt} \Bigr>=\sum_{\ell,m,n}
{\Bigl|Z_{\ell m\omega_n}\Bigr|^2\over4\pi\omega_n^2}
\equiv \sum_{\ell,m,n}\Bigl( {dE \over dt} \Bigr)_{\ell mn}.\eqn\flu
$$
In the same way, the  angular momentum flux is given by
$$
\Bigl< {dJ_z \over dt} \Bigr>=\sum_{\ell,m,n}
{m\Bigl| Z_{\ell m\omega_n}\Bigr|^2\over4\pi\omega_n^3}
\equiv \sum_{\ell,m,n}\Bigl( {dJ_z \over dt} \Bigr)_{\ell mn}
=\sum_{\ell,m,n}{m\over\omega_n}\Bigl( {dE \over dt} \Bigr)_{\ell mn}\,.
\eqn\jflux
$$
As clear from Eq.\Zform,
since we take the square of $|Z_{\ell m\omega_n}|$
we only need $n=0$, $\pm1$ modes for the present purpose.
Further, since
$\omega_n=m\Omega_{\varphi}+n\Omega_{\theta}$, we only need
 $\Omega_\theta$ at the 0th order of $y$. The PNE of $\Omega_\varphi$
and $\Omega_\theta$ are given in Eq.\omeeq.
In order to express the post-Newtonian corrections to the luminosity,
we define $\eta_{\ell mn}$ as
$$
\left( {dE \over dt} \right)_{\ell mn}
\equiv{1\over2}\left({dE\over dt}\right)_N \eta_{\ell mn}\,,
\eqno\eq
$$
where $(dE/dt)_N$ is the Newtonian quadrupole luminosity:
$$
\left({dE\over dt}\right)_N={32\mu^2M^3\over5r_0^5}
={32\over5}\left({\mu\over M}\right)^2v^{10}.
\eqn\Enewton
$$
We note that because the mode indices $(\ell,m)$ here are those
associated with the spheroidal harmonics, as clear from Eq.\heq,
they do not correspond to the usual spherical mode indices.
This point should be kept in mind when one attempts to interpret
the PN corrections to $\eta_{\ell mn}$ in the language of the
standard PN approach.

For $\ell=2$, the results are as follows.
If $|m+n|>2$ or $m+n=0$,
$\eta_{\ell mn}$ becomes $O(v^k)$ $(k>5)$.
The remaining $\eta_{\ell mn}$ which contribute to the luminosity to
$O(v^5)$ are given by
\def\pmz{\hphantom{\pm}0}
$$\eqalignno{
\eta_{2\pm 2\pmz}=
&1-{107 \over 21}v^2+4\pi v^3-6q v^3+{4784 \over 1323}v^4
+2q^2v^4-{428\pi \over 21}v^5+{4216 \over 189}qv^5
&\cr
&+y \Bigl( -1+{170 \over 21}v^2-4\pi v^3+15qv^3-{4784 \over 1323}v^4
  -11q^2v^4
&\cr
&\qquad+{428\pi \over 21}v^5-{13186\over 189}qv^5 \Bigr),
&\cr
\eta_{2\pm2\mp1}=
&y\Bigl( {1 \over 36}v^2 - {17 \over 504}v^4
+{\pi \over 18}v^5+{17 \over 1134}q v^5 \Bigr),
&\cr
\eta_{2\pm1\pmz}
=&{1 \over 36}v^2-{1 \over 12}q v^3-{17 \over 504}v^4
+{1 \over 16}q^2v^4+{\pi \over 18}v^5-{793 \over 9072}q v^5
&\cr
&+y \Bigl( -{5 \over 72}v^2 +{1 \over 8}q v^3 +{85 \over 1008}v^4
-{1 \over 32}q^2v^4-{5\pi \over 36}v^5+{13931 \over 18144}qv^5 \Bigr),
&\eqnalign\etatwo\cr
\eta_{2\pm1\pm1}=
&y\Bigl( 1- {170 \over 21}v^2 +4\pi v^3 -12q v^3
+{4784 \over 1323}v^4+{11 \over 2}q^2v^4
&\cr
&\qquad-{428\pi \over 21}v^5+{11078 \over 189}q v^5 \Bigr),
&\cr
\eta_{2\pmz\pm1}=
&y\Bigl(  {1 \over 24}v^2  -{1 \over 12}q v^3
-{17 \over 336}v^4+{1 \over 24}q^2v^4
+{\pi \over 12}v^5-{745 \over 1008}qv^5 \Bigr).
&\cr}
$$
Putting together the above results,
we obtain $(dE/dt)_\ell \equiv \sum_{mn} (dE/dt)_{\ell mn}$
for $\ell=2$ as
$$\eqalign{
\Bigl( {dE \over dt} \Bigr)_2=
&\left({dE\over dt}\right)_N
\Bigl\{ 1 -{1277 \over 252}v^2 +4\pi v^3
-{73 \over 12}qv^3\Bigl(1-{y \over 2}\Bigr)
+{37915 \over 10584}v^4 \cr
&+{33 \over 16}q^2v^4 -{527 \over 96}q^2v^4y
-{2561\pi \over 126}v^5+{201575 \over 9072}qv^5\Bigl(1-{y\over2}\Bigr)
\Bigr\}.\cr}
\eqn\Etwo
$$

For $\ell=3$, the non-trivial $\eta_{\ell mn}$ are given by
$$\eqalignno{
\eta_{3\pm3\pmz}=
&{1215 \over 896}v^2 -{1215 \over 112}v^4
+{3645\pi \over 448}v^5 -{1215 \over 112}qv^5
&\cr
&+y \Bigl( -{3645 \over 1792}v^2 +{3645 \over 224}v^4
-{10935\pi \over 896}v^5 +{3645 \over 112}qv^5 \Bigr),
&\cr
\eta_{3\pm3\mp1}=&{5 \over 42}v^4 y,
&\cr
\eta_{3\pm2\pmz}=&{5 \over 63}v^4 -{40 \over 189}qv^5
+y\Bigl( -{20 \over 63}v^4+{100 \over 189}qv^5\Bigr),
&\cr
\eta_{3\pm2\pm1}=&y\Bigl( {3645 \over 1792}v^2
-{3645 \over 224}v^4
+{10935\pi \over 896}v^5-{6075 \over 224}qv^5\Bigr),
&\cr
\eta_{3\pm2\mp1}=&y\Bigl( {5 \over 16128}v^2
-{5 \over 3024}v^4
+{5\pi \over 8064}v^5+{25 \over 18144}qv^5\Bigr),
&\eqnalign\etathr\cr
\eta_{3\pm1\pmz}=&{1 \over 8064}v^2 -{1 \over 1512}v^4
+{\pi \over 4032}v^5 -{17 \over 9072}qv^5
&\cr
&+y \Bigl( -{11 \over 16128}v^2 +{11 \over 3024}v^4
-{11\pi \over 8064}v^5 +{95 \over 9072}qv^5 \Bigr),
&\cr
\eta_{3\pm1\pm1}=
&y\Bigl( {25 \over 126}v^4 -{80 \over 189}qv^5\Bigr),
&\cr
\eta_{3\pmz\pm1}=&y\Bigl( {1 \over 2688}v^2 -{1 \over 504}v^4
+{\pi \over 1344}v^5-{11 \over 1008}qv^5 \Bigr).
&\cr}
$$
The other $\eta_{\ell mn}$ are of $O(v^6)$ or higher.
Then we obtain
$$
\Bigl( {dE \over dt} \Bigr)_3=
\left({dE\over dt}\right)_N
\Bigl\{  {1367 \over 1008}v^2 -{32567 \over 3024}v^4
+{16403\pi \over 2016}v^5-{896 \over 81}qv^5\Bigl(1-{y \over 2}\Bigr)
 \Bigr\}.
\eqn\Ethr
$$

For $\ell=4$, we have
$$\eqalign{
&\eta_{4\pm4\pmz}={1280 \over 567}v^4(1-2y),\cr
&\eta_{4\pm3\pm1}={2560 \over 567}v^4 y, \cr
&\eta_{4\pm3\mp1}={5 \over 1134}v^4 y, \cr
&\eta_{4\pm2\pmz}={5 \over 3969}v^4(1-8y), \cr
&\eta_{4\pm1\pm1}={5 \over 882}v^4 y, \cr}\eqn\etafour
$$
and the others are of $O(v^6)$ or higher.
Hence we obtain
$$
\Bigl( {dE \over dt} \Bigr)_4=
\left({dE\over dt}\right)_N
\times {8965 \over 3969}v^4. \eqn\Efour
$$

Finally, gathering all the above results, the total energy flux
up to $O(v^5)$ is found to be
$$\eqalign{
\left\langle {dE \over dt} \right\rangle =
\left({dE\over dt}\right)_N
&\biggl( 1 -{1247 \over 336}v^2+4\pi v^3
 -{73 \over 12}qv^3\left(1-{y\over2}\right)
 -{44711 \over 9072}v^4 \cr
&+{33 \over 16}q^2v^4-{527 \over 96}q^2v^4 y -{8191 \pi \over 672}v^5
+{3749 \over 336}qv^5\left(1-{y\over2}\right) \biggr).\cr}
\eqn\Etot
$$
The terms without $q$ agree with those derived in Ref.[\tagosas],
and the term $-73qv^3/12$ also agrees with the previous
results\rlap.\refmark{\Kidder,\poisa}\footnote{\dag}{
Note that Poisson\refmark{\poisa}
defines the PN expansion parameter as
$v'\equiv(M\Omega_\varphi)^{1/3}$, which is related to our
$v\equiv(M/r_0)^{1/2}$ as $v'=v\bigl(1-qv^3/3+O(v^6)\bigr)$ for $y=0$.
Consequently, his Newtonian quadrupole luminosity
differs from ours by a factor
$(v'/v)^{10}=\bigl(1-10qv^3/3+O(v^4)\bigr)$. This explains
the apparent difference between his result $-11qv^3/4$ and ours.}\

\ From Eq.\jflux, the avaraged angular momentum fluxes for $\ell=2$, 3
and 4 are calculated to give
$$
\eqalignno{
\left({dJ_z\over dt}\right)_2
=&\left({dJ_z\over dt}\right)_N
 \biggl[1-{y\over 2}
-{1277\over 252}v^2\left(1-{y\over2}\right)
+4\pi v^3\left(1-{y\over 2}\right)
&\cr
&-qv^3\left({61\over 12}-{61\over 8}y\right)
+{37915\over 10584}v^4\left(1-{y\over2}\right)
+q^2v^4\left({33 \over 16}-{229 \over 32}y\right)
&\cr
&-\pi v^5{2561 \over 126}\left(1-{y\over2}\right)
+qv^5\left({22229 \over 1296}-{27809\over864}y\right)
\biggr],
&\eq\cr
\left({dJ_z\over dt}\right)_3
=&\left({dJ_z\over dt}\right)_N
\biggl[{1367\over 1008}v^2\left(1-{y \over 2}\right)
-{32567 \over 3024}v^4\left(1-{y\over2}\right)
&\cr
&
+\pi v^5{16403\over 2016}\left(1-{y\over 2}\right)
-qv^5\left({88049 \over 9072}-{9817\over756}y\right)
\biggr],
&\eq\cr
\left({dJ_z\over dt}\right)_4
=&\left({dJ_z\over dt}\right)_N
\biggl[{8965\over 3969}v^4\left(1-{y\over 2}\right)\biggr],
&\eq\cr}
$$
where $(dJ_z/dt)_N$ is defined to be
$$
\left({dJ_z\over dt}\right)_N
={32\mu^2M^{5/2}\over5r_0^{7/2}}
={32\over5}\left({\mu\over M}\right)^2Mv^7.
\eqn\Jnewton
$$
Total flux of the angular momentum is then given by
$$
\eqalign{
\left\langle{dJ_z\over dt}\right\rangle
=&\left({dJ_z\over dt}\right)_N
\biggl[
\left(1-{y\over2}\right)-{1247\over336}v^2\left(1-{y\over2}\right)
+4\pi v^3\left(1-{y\over2}\right)
\cr
&-{61\over12}qv^3\left(1-{y\over2}\right)
-{44711\over9072}v^4\left(1-{y\over2}\right)
+q^2v^4\left({33\over16}-{229\over32}y\right)
\cr
&-{8191\over672}\pi v^5\left(1-{y\over2}\right)
+qv^5\left({417\over56}-{4301\over224}y\right)
\biggr].\cr}
\eqn\Jtot
$$
We note that the result is proportional to $(1-y/2)$
in the limit $q\rightarrow0$. This is simply because
the orbital plane is slightly tilted from the equitorial plane
by an angle $\theta_i\sim y^{1/2}$, hence
$dJ_z/dt\sim (dJ_{tot}/dt)\cos\theta_i$.

\ From the above results, we find the following features
of the gravitational wave luminosity:

\noindent
1) As argued in \S 2.1,
the quadratic terms in $q(=a/M)$ appear at the $v^4$ order, and the linear
terms in $q$ appear at $v^3$ and $v^5$ orders.

\noindent
2) The coefficients of $qv^3$ and $qv^5$ in $(dE/dt)_2$ and
 $(dE/dt)_3$ (and hence in $(dE/dt)_{tot}$) are proportional to
$1-y/2$. Since $1-y/2\sim\cos\theta_i$,
these terms may be regarded as proportional to the inner product
$\vec S\cdot\vec L$ of
the spin angular momentum $\vec S$ and the orbital angular momentum
$\vec L$. With this interpretation,
our result at the $v^3$ order is consistent with the PN calculation of the
spin-orbit terms by Kidder et al.\refmark{\Kidder}
as well as with the numerical result of perturbative calculations
by Shibata\rlap.\refmark\shiba\

\noindent
3) Contrary to the feature 2), the coefficient of the $q^2v^4$
term in $(dE/dt)_2$ does not seem to be expressible as a simple function
of $\cos\theta_i$. We suspect that a major part of it is attributable
to the quadrupolar gravitational field around the Kerr black hole
which modifies the particle orbit. In fact, for $y=0$,
the $2q^2v^4$ term of $\eta_{2\pm20}$ in Eq.\etatwo\ can be explained
in terms of the Newtonian quadrupole formula
 as the contribution from the quadrupole moment
of the Kerr black hole\rlap.\footnote{\star}{This was first pointed to
us by E. Poisson.}\
However, the $1/16\,q^2v^4$ term in $\eta_{2\pm10}$ cannot be explained
 in this way. An inspection of the
expanded form of the $\ell=2$ ingoing-wave Teukolsky function given in
Eq.\elltwo\ reveals that the $q^2z^2$ term at $O(\epsilon^2)$ is
proportional to $m^2-4$, hence it vanishes
for $m=\pm2$ while it remains finite for $m=\pm1$. Since this term will
contribute to the $q^2v^4$ terms in the luminosity, we may interpret the
$q^2v^4$ term in $\eta_{2\pm10}$ as due to the curvature
scattering in the near zone field. Incidentally, this suggests that
the coincidence of the coefficient 2 of the $q^2v^4$ term in
$\eta_{2\pm20}$ with the Newtonian calculation is rather accidental;
naively we would expect the same curvature scattering effect to give
rise to some additional contribution to the $q^2v^4$ term in
 $\eta_{2\pm20}$. In any case, our result
suggests the existence of a new type of spin-dependent terms
in the energy flux when a PN analysis beyond the
present level is carried out.

\chapter{Implications to coalescing compact binaries}

In this section, we discuss the effects of PN terms in the
luminosity to the orbital evolution of inspiraling binaries
 composed of neutron stars (NSs) and/or black holes (BHs).
Although our results are valid only in the test particle limit, we ignore
this fact in the following.

Since we are interested in the orbits off the equatorial plane,
we must consider the evolution of
$C$ as well as $r$ of the particle as it radiates gravitational waves.
Although not proved in any sense, let us we assume that the orbit
remains quasi-circular, \ie, the radius of the orbit is approximately
constant for many orbital periods. We then would like to see if the
inclination angle $\theta_i(=y^{1/2})$ changes in time as the orbit shrinks.
Since the test particle orbit in this case is characterized by two of the
four approximate constants of motion, $r$, $C$ (or $y$), $E$ and $l_z$,
we can estimate the change of $y$ with respect to the change of $r$
by equating the energy and angular momentum luminosities of the
gravitational waves with those lost by the particle,
$$
\eqalign{
-\VEV{{dE\over dt}}
=&{\partial E(r,y)\over\partial r}{dr\over dt}
                      +{\partial E(r,y)\over\partial y}{dy \over dt}\,,
\cr
-\VEV{{dJ_z\over dt}}
=&{\partial l_z(r,y)\over\partial r}{dr\over dt}
                      +{\partial l_z(r,y)\over\partial y}{dy \over dt}\,.
\cr}
\eqn\EJval
$$
To the leading order in $v$ and $y$, this gives
$$
{d\ln y\over d\ln r}=-{61\over24}q\left({M\over r}\right)^{3/2}.
\eqn\dydr
$$
Thus $y$ changes only by a small amount
during the entire inspiraling stage until $r\lsim10M$ even for $q=1$.
Hence the approximation $y=const.$ throughout the evolution of the orbit
will be good if the orbit remains quasi-circular.
Furthermore, by adopting a radiation reaction formula
 which is valid at least in the Newtonian limit,
 it has been numerically found by Shibata\refmark{\shiba} that
the evolution of the inclination angle is small at least
in the low frequency region, $r/M \gsim 30$.
This result is consistent with the assumption of quasi-circular orbits.
Thus we assume $y=const.$ in the following.

Then the total cycle $N(r_i,r_f)$ of the phase of gravitational waves
from an inspiraling compact binary from
 $r=r_i$ ($t=t_i$) to $r=r_f$ ($t=t_f$) is
$$
N\equiv \int^{t_f}_{t_i} fdt
={1 \over \pi} \int^{r_i}_{r_f}dr \Omega_{\varphi}
{dE/dr \over |dE/dt|},\eqno\eq
$$
where $f$ is the frequency of the wave.
Expanding $\Omega_{\varphi}$, $dE/dt$, and $dE/dr$
with respect to $v=(M/r)^{1/2}$, $N$ is expressed as
$$
N={5\over 64\pi}{M\over \mu}\int^{r_i}_{r_f}{dr r^{3/2}\over M^{5/2}}
{{\sum^\infty_{k=0}b_k(q)(M/r)^{k/2}
\sum^{\infty}_{k=0}c_k(q)(M/r)^{k/2}}
\over \sum^{\infty}_{k=0}d_k(q)(M/r)^{k/2}},
\eqno\eq
$$
where the series forms in the denominator and numerator represent
the PN corrections to the $\Omega_{\varphi}$, $dE/dt$,
and $dE/dr$, that is
$$\eqalign{
&\sum^\infty_{k=0} b_k(q)(M/r)^{k/2}
={\Omega_{\varphi}\over(\Omega_{\varphi})_N}\,,\cr
&\sum^\infty_{k=0}c_k(q)(M/r)^{k/2}
={(dE/dr)\over(dE/dr)_N}\,,\cr
&\sum^\infty_{k=0}d_k(q)(M/r)^{k/2}
={(dE/dt)\over(dE/dt)_N}\,,\cr}
\eqno\eq
$$
and the argument $q (=a/M)$ is given to the coefficients $b_k$, $c_k$
and $d_k$ to emphasize the $q$-dependence of the PN corrections.
To the PN order we consider in this paper,
the PN expansions of $\Omega_\varphi$ and $dE/dt$ are given in
Eqs.\omeeq\ and \Etot, respectively. For completeness, here we show the
PN expansion of $dE/dr$:
$$
\eqalign{
 {dE\over dr}={\mu M\over2r^2}&\biggl[
 1-{3M\over2r}+5q(1-{y\over2})\left({M\over r}\right)^{3/2}
\cr
&-\left({81\over8}+3q^2(1-y)\right)\left({M\over r}\right)^{2}
-{21\over4}q\left({M\over r}\right)^{5/2}\biggr].
\cr}
\eqn\dEdr
$$

Since the effect of PN corrections in the case
of $q=0$ has been already studied\rlap,\refmark\tagosas\
here we examine only the effect due to non-vanishing $q$.
For this purpose, we introduce the quantity $\Delta N^{(n)}$ defined by
$$
\eqalign{
\Delta N^{(n)}(q)={5\over 64\pi}{M\over \mu}
&\Biggl[
\int^{r_i}_{r_f}{dr r^{3/2}\over M^{5/2}}
{{\sum^n_{k=0}b_k(q)(M/r)^{k/2}\sum^{n}_{k=0}c_k(q)(M/r)^{k/2}}
\over \sum^{n}_{k=0}d_k(q)(M/r)^{k/2}}
\cr
&-\int^{r_i}_{r_f}dr r^{3/2}\biggl\{
{{\sum^{n-1}_{k=0}b_k(q)(M/r)^{k/2}+b_n(0)(M/r)^{n/2}}
\over \sum^{n-1}_{k=0}d_k(q)(M/r)^{k/2}+d_n(0)(M/r)^{n/2}}\cr
&\qquad
\times\Bigl(
\sum^{n-1}_{k=0}c_k(q)(M/r)^{k/2}+c_n(0)(M/r)^{n/2}\Bigr)\biggr\}
\Biggr].\cr}
\eqn\DelN
$$
This describes the effect of $q$ corrections at the P$^{n/2}$N order.

First we consider a NS-NS binary of equal mass $M=1.4M_{\odot}$ ($M/\mu=4$)
as a typical example. The future laser interferometric
gravitational wave detectors
such as LIGO\refmark{\ligo} have good sensitivity in the frequency
band between $10$ to $1000$Hz, so we choose $r_i=175M$ and
$r_f=8M$.
A NS of mass $1.4 M_{\odot}$ and radius $R\simeq10$km
has $q\simeq 0.4/P_{ms}$, where $P_{ms}$ is the period of rotation
in units of a mili-second.
In this case, the phase $N$ is accumulated at relatively large radii
$r/M\gsim100$. Hence the convergence of the PN series in the numerators
and denominators of the integrands in Eq.\DelN\ is good enough so that
we may expand them further to obtain the approximate formula,
$$
\eqalign{
\Delta N^{(n)}(q)\approx&{5\over64\pi}{M\over\mu}
\int_{r_f}^{r_i}{dr\over M}\left({M\over r}\right)^{n-3/2}
\cr
&\times
\left\{\bigl(b_n(q)+c_n(q)-d_n(q)\bigr)-\bigl(b_n(0)+c_n(0)-d_n(0)\bigr)
\right\}.
\cr}
\eqn\Nexpand
$$
This gives
$$
\eqalign{
\Delta N^{(3)}&\sim \left({70\over P_{ms}}\right)(1-0.4y)\,,\cr
\Delta N^{(4)}&\sim -\left({2\over P_{ms}}\right)^2(1-1.5y)\,,\cr
\Delta N^{(5)}&\sim \left({9\over P_{ms}}\right)(1-0.3y)\,.\cr
}\eqn\NSNS
$$
Although the $y$-corrections in the above are valid only for $y\ll1$,
we expect them to be qualitatively valid even for $y=O(1)$.
As mentioned previously, the corrections $\Delta N^{(3)}$ and
$\Delta N^{(5)}$ are due to the spin-orbit coupling, hence replacing
$y$ with $2(1-\cos\theta_i)$ in the above formula will give a reasonable
estimate in the qualitative sense.
As for $\Delta N^{(4)}$, although we have not been able to specify the
physical meaning of it with certainty, at least we may say
 that the dominant contribution
comes from the quadrupole moment of the gravitational field around the
Kerr black hole, as discussed at the end of the previous section.
Hence we may also expect the replacement $y\rightarrow2(1-\cos\theta_i)$
to be approximately correct.
Thus the inclination angle directly affects the values of these
phase corrections whenever they become important.

We know three binary pulsars
in our Galaxy which will merge within a Hubble time,
 PSR1913+16\rlap,\refmark\Tay\
PSR2127+11C\refmark{\Ander} and PSR1534+12\rlap.\refmark\Wol\
Hence these may be regarded as
a typical target of the gravitational wave detectors.
Their rotation periods are $P_{ms}=59.0$, $30.5$ and $37.9$, respectively.
If we also assume these values as typical,
we have $\Delta N^{(3)}>1$
as has been discussed previously\rlap,\refmark{\shi,\shiba}\
 while $\Delta N^{(4)}$ and $\Delta N^{(5)}$ are small.
However, we know there are several pulsars with $P_{ms}\lsim 2$ in our
Galaxy\rlap,\refmark\Puls\ for which both $\Delta N^{(4)}$ and
$\Delta N^{(5)}$ exceed unity.
Hence  it will be safe to construct templates which take
account of the ${\rm P^{2}N}$ and ${\rm P^{5/2}N}$ spin terms.

Note however that a main contribution to the correction $\Delta N^{(4)}$
is due to the quadrupole moment of the gravitatinal field,
the value of which reflects the peculiarity of the Kerr black hole.
Concerning this point, Bildsten and Cutler\refmark{\bild} considered
the quadrupole moment of the gravitational field induced by the
quadrupolar deformation of a NS due to its rotation and
evaluated the phase correction for a realistic NS model as
$$
\Delta N^{(4)} \sim -\Bigl({5 \over P_{ms}}\Bigr)^2.\eqno\eq.
$$
This result is somewhat larger than our estimate.
That is, the effect of the quadrupole moment of the gravitational
field due to a spinning NS is larger than that due to a spinning BH
for the same dimensionless spin parameter $q$.
This suggests that if one body of a compact binary is very rapidly
rotating so that we are able to measure $\Delta N^{(4)}$
by the matched filter technique, then together with other terms which
carry information of
the orbital parameters such as the spin-orbit terms,
it will be possible to distinguish a BH from a NS
even if the BH has mass $M_{BH} \sim 1.5M_{\odot}$.

Next, we consider a BH-NS binary composed of
$10M_{\odot}$ BH and $1.4M_{\odot}$ NS ($M/\mu=9.28$).
Our result has more direct applicability to this case.
For simplicity,
we set $r_i=68M$ and $r_f=6M$ irrespective of $q$ and $y$.
In this case, the estimation of $\Delta N^{(n)}$ in terms of the
approximate formula \Nexpand\ will not be a good approximation.
 Instead we must use the original formula for
$\Delta N^{(n)}$, Eq.\DelN, as it is.
This is because
the phase $N$ in the present case is accumulated at smaller $r/M$
than in the case of a NS-NS binary, hence
the convergence of the PN expansion becomes slow.
As a consequence it is not possible to derive approximate formulas
 for $\Delta N^{(n)}$ as simple as Eq.\NSNS. Here we only quote
the critical value of the spin parameter $q$ above which each
correction $\Delta N^{(n)}$ $(n=3,4,5)$ exceeds unity.
We find $\Delta N^{(3)}\gsim1$ for $q\gsim0.01$. Hence the correction at
this order is important even for a very slowly rotating black hole.
As for $\Delta N^{(4)}$ and
$\Delta N^{(5)}$, they become larger than unity if $q\gsim0.2$.
This indicates that yet higher order PN corrections will be important
if the BH is rapidly rotating ($q\sim1$).

Summarizing the above analyses, we obtain the following conclusions:

\noindent
1) The spin-orbit coupling term at the P$^{5/2}$N order
is important for the evolution of NS-NS binaries with $P_{ms}\lsim2$
and BH-NS binaries with $q\gsim0.2$. Since the rotation of $P_{ms}\sim1$
would be the fastest possible period that a NS could have, inclusion of
the phase corrections up through the P$^{5/2}$N order seems to be enough
for NS-NS binaries. On the other hand, the terms higher
than the ${\rm P^{5/2}N}$ order are likely to be important for BH-NS
binaries since the rotation of $q\gsim0.2$ for a black hole seems quite
possible.

\noindent
2) The $q^2$ terms at the P$^2$N order becomes important for
BHs with $q\gsim0.2$ or NSs with $P_{ms}\lsim2$. However, the latter
value is based on our formula which is valid only for a rotating BH.
An estimate base on a realistic NS model gives
$P_{ms}\lsim5$\rlap.\refmark\bild\
Thus it will be possible to distinguish a small mass BH from a NS if the
phase corrections to P$^2$N order can be detected by matched filtering.

\noindent
3) At any order of PN corrections,
the effect of a finite inclination angle to the number of the
phase cycles must be taken into account
whenever the spin terms become important.

\chapter{Summary}

In this paper, we have performed a post-Newtonian
calculation of the gravitational waves from a particle of mass
$\mu$ orbiting around a rotating black hole of mass $M$ $(\mu\ll M)$.
We have considered the orbit of a constant coordinate radius $r=r_0$
but with small inclination angle $\theta_i\sim y^{1/2}$, where
$y$ is a non-dimensional parameter proportional to the Carter
constant of the orbit.

We have formulated the post-Newtonian expansion of the Teukolsky
equation and its source term in terms of a small expansion parameter
$v=(M/r_0)^{1/2}$ accurate up through $O(v^5)$ (P$^{5/2}$N order).
We have not directly dealt with the Teukolsky equation
but first formulated a method to obtain the homogeneous solution to
the Sasaki-Nakamura equation, which is a generalization of the
Regge-Wheeler equation for the Schwarzschild black hole,
by expanding it in powers of $\epsilon=2M\omega$,
where $\omega$ is the frequency of a gravitational wave.
In particular, to $O(\epsilon)$,
we have obtained the ingoing-wave radial
functions for arbitrary spherical indices $(\ell,m)$
in closed analytical form.
Then we have obtained all the necessary radial functions to the required
accuracy and transformed them to the corresponding Teukolsky radial
functions, which have been used to construct the Green function.
Further, we have formulated the post-Newtonian expansion of
the source term for circular orbits with small inclination angle.
Assuming $y\ll1$, we have analytically solved the geodesics of a particle
accurate to $O(y)$ and obtained the source term to the required
accuracy. We have used these results to integrate the Teukolsky
equation and derived the formulas for the gravitational
energy and angular momentum fluxes which are correct to $O(v^5)$
and to $O(y)$.

Based on thus obtained luminosity formula,
we have estimated the accumulated phase $N$
of gravitational waves from inspiraling binaries,
assuming the orbit remains quasi-circular.
Specifically we have considered a NS-NS binary
of equal mass $1.4M_{\odot}$ and a BH-NS binary
of masses $10M_{\odot}$ and $1.4M_{\odot}$, which will be
typical targets of the near-future laser interferometric gravitational wave
detectors. We have found that if the rotation of a neutron star is
moderate, say $P\gsim 20\,{\rm ms}$, only the phase correction at its
leading P$^{3/2}$N order will be important.
However we have also found that if one body of a binary
is a rapidly rotating NS ($P\lsim 2\,{\rm ms}$) or
a rotating BH ($q=J_{BH}/M^2 \gsim 0.2$),
the phase correction of $\Delta N>1$ will be caused by
the spin terms at ${\rm P^{2}N}$ and ${\rm P^{5/2}N}$ orders.
Furthermore if one body is
a rapidly rotating BH ($q\sim 1$), it is expected
that the higher order corrections such as P$^3$N ones become important.
In all these cases, when the phase correction at a certain PN order
become significant, that due to a non-vanishing inclination angle at the
same PN order becomes equally important.

The above conclusions imply that it is desirable to evaluate yet
higher order PN spin corrections
to the gravitational wave luminosity.
As for the inclination of the orbit, since we expect
the expansion in powers of $y$ to be valid for
$y \lsim 1$ $(\theta_i \lsim \pi/4)$,
it will be meaningful and useful to calculate
the higher order corrections in $y$ along with higher order PN calculations.
These problems are left for future work.

\bigskip
\leftline{\bf Acknowledgements}

We thank E. Poisson for useful discussions.
H.T. and T.T. thank Prof. H. Sato for continuous encouragement.
This work was partly supported by the Japanese Grant-in-Aid Scientific
Research of the Ministry of Education, Science
and Culture, No. 04234104 and No. 06740343.

\Appendix{A}

In this Appendix we show the potential functions $F$ and $U$
of the SN equation \sneq. Details of the derivation are
given in Ref.[\Sas].

The function $F(r)$ is given by
$$
F(r)={\eta_{,r} \over \eta}{\Delta \over r^2+a^2},
\eqn\Fdef
$$
where
$$
\eta=c_0+c_1/r+c_2/r^2+c_3/r^3+c_4/r^4,
\eqn\etadef
$$
with
$$\eqalign{
&c_0=-12i\omega M+\lambda(\lambda+2)-12a \omega(a\omega-m),\cr
&c_1=8ia[3a\omega-\lambda(a\omega-m)],\cr
&c_2=-24ia M(a\omega-m)+12a^2[1-2(a\omega-m)^2 ],\cr
&c_3=24ia^3(a\omega-m)-24Ma^2,\cr
&c_4=12a^4.\cr}\eqn\ceq
$$
The function $U(r)$ is given by
$$
U(r)={\Delta U_1\over(r^2+a^2)^2}+G^2+{\Delta G_{,r}\over r^2+a^2}-FG,
\eqn\Udef
$$
where
$$\eqalign{
&G=-{2(r-M) \over r^2+a^2}+{r\Delta \over (r^2+a^2)^2},\cr
&U_1=V+{\Delta^2 \over \beta}
\Bigl[\Bigl(2\alpha+{\beta_{,r}  \over \Delta}\Bigr)_{,r}
-{\eta_{,r} \over \eta}
\Bigl(\alpha +{\beta_{,r} \over \Delta}\Bigr)\Bigr],\cr
&\alpha=-i{K \beta \over \Delta^2}+3iK_{,r}
+\lambda+{6\Delta \over r^2},\cr
&\beta=2\Delta\Bigl(-iK+r-M-{2\Delta \over r}\Bigr).\cr}\eqn\ur
$$

\Appendix{B}

In this Appendix we show the $A$'s in Eq.\sour.
$$\eqalign{
A_{n\,n\,0}&={2 \over \sqrt{2\pi}\Delta^2}\int_0^{\Delta t}dt
e^{i\omega t-im\varphi(t)}C_{n\,n}\rho^{-2}{\overline \rho}^{-1}
L_1^+\{\rho^{-4}L_2^+(\rho^3 S)\},\cr
A_{{\overline m}\,n\,0}&=-{2 \over \sqrt{\pi}\Delta}\int_0^{\Delta t}dt
e^{i\omega t-im\varphi(t)} C_{{\overline m}\,n}\rho^{-3}
\Bigl[\left(L_2^+S\right)
\Bigl({iK \over \Delta}+\rho+{\overline \rho}\Bigr)
\cr
&\qquad-a\sin\theta S {K \over \Delta}({\overline \rho}-\rho)\Bigr],\cr
A_{{\overline m}\,{\overline m}\,0}
&={1 \over \sqrt{2\pi}}\int_0^{\Delta t}dt
e^{i\omega t-im\varphi(t)} \rho^{-3}{\overline \rho}
C_{{\overline m}\,{\overline m}}S\Bigl[
-i\Bigl({K \over \Delta}\Bigr)_{,r}-{K^2 \over \Delta^2}+
2i\rho {K \over \Delta}\Bigr],\cr
A_{{\overline m}\,n\,1}&=-{2\over \sqrt{\pi}\Delta }\int_0^{\Delta t}dt
e^{i\omega t-im\varphi(t)} \rho^{-3}
C_{{\overline m}\,n}
[L_2^+S+ia\sin\theta({\overline \rho}-\rho)S],\cr
A_{{\overline m}\,{\overline m}\,1}
&={2 \over \sqrt{2\pi}}\int_0^{\Delta t}dt
e^{i\omega t-im\varphi(t)} \rho^{-3}{\overline \rho}
C_{{\overline m}\,{\overline m}}S\Bigl(i{K \over \Delta}+\rho\Bigr),\cr
A_{{\overline m}\,{\overline m}\,2}
&={1\over \sqrt{2\pi}}\int_0^{\Delta t}dt
e^{i\omega t-im\varphi(t)} \rho^{-3}{\overline \rho}
C_{{\overline m}\,{\overline m}}S,\cr
} \eqno\eq
$$
where $S$ denotes $_{-2}S_{\ell m}^{a\omega}$.

\Appendix{C}

In this Appendix, we show that the asymptotic form of
$X^{in}_{\ell m\omega}$ at $\epsilon\ll z\ll 1$ has the form,
$$
X^{in}_{\ell m\omega}=z^{l+1}[O(1)+\epsilon O(z^{-1})
+\epsilon^2 O(z^{-2})+\epsilon^3 O(z^{-3})+\ldots. \eqn\xasy
$$

\ From Eqs.$\sneq$ and $\xixi$, the equation for $\xi_{\ell m}$
becomes
$$
\eqalign{
&\left[{\Delta\over r^2}{{d^2}\over dr^2}
+C_1{d\over dr}+C_2\right]\xi_{\ell m}=0;
\crr
&~C_1={{\Delta_{,r}-\Delta F_1-2i\Delta\phi_{,r}}\over r^2}, \cr
&~C_2={1\over r^2}
\left(-U_1+2-{\Delta_{,r}^2\over\Delta}-\Delta\phi_{,r}^2-\Delta_{,r}F_1
+i\Delta\phi_{,r}F_1-2iM\omega\right),
\cr}
\eqn\eqxi
$$
where
$F_1=\eta_{,r}/\eta$, and $\phi$, $\eta$ and $U_1$ are given in
Eqs.\phase, \etadef\ and \ur, respectively.

Let us now examine the behaviors of the coefficients $C_1$ and $C_2$.
By dimensional consideration, they must be of the forms,
$$
C_k={1\over r^k}\,f_k(2M/r, \omega r)\quad(k=1,2),
\eqn\Cform
$$
where $f_k(y,z)$ are dimensionless functions of their arguments.
We then easily see that $f_k(y,z)$ are regular
at $y=0$, since both $C_1r$ and $C_2r^2$ have well-defined limits as
$M\rightarrow0$. Furhtermore, by examining the behaviors of
$C_1$ and $C_2$ as $\omega\rightarrow0$, we find they are also
regular in this limit. Hence $f_k(y,z)$ are regular at $z=0$ as well.
Thus $f_k$ may be expanded as
$$
f_k(y,z)=\sum_{n=0}^\infty f_k^{(n)}(z)y^n,
\eqn\fexpand
$$
where $f_k^{(n)}(z)$ are regular at $z=0$,
hence may be futher expanded as
$$
 f_k^{(n)}(z)=\sum_{m=0}^\infty f_k^{(n,m)}z^m.
\eqno\eq
$$
Note that $f_1^{(0)}(z)=2$ and $f_2^{(0)}(z)=-\ell(\ell+1)+z^2$,
which are the coefficients appearing in the lowest
order differential operator $L^{(0)}$ given by Eq.\Lzero.

Taking the above consideration into account,
scaling $r$ to $z=\omega r$ in Eq.\eqxi, and noting that
$y=2M/r=\epsilon/z$, we have
$$
\eqalign{
\Biggl[L^{(0)}+&\left(-{\epsilon\over z}
+{q^2\over4}{\epsilon^2\over z^2}\right){d^2\over dz^2}
\cr
&+\left(\sum^{\infty}_{n=1}f_1^{(n)}(z){\epsilon^n\over z^n}\right)
{1\over z}{d\over dz}
+\left(\sum^{\infty}_{n=1}f_2^{(n)}(z){\epsilon^n\over z^n}\right)
\Biggr]\xi_{\ell m}=0. \cr} \eqn\eqex
$$
Accordingly, if we expand $\xi_{\ell m}$ as
$\xi_{\ell m}(\epsilon;z)
=\sum_{n=0}^{\infty}\epsilon^n \xi_{\ell m}^{(n)}(z)$
and set $\xi_{\ell m}^{(0)}(z)=j_l\sim z^\ell$,
we can easily show recursively that the
asymptotic behavior of $\xi_{\ell m}^{(n)}$ at $z\ll 1$ is
$$
\xi_{\ell m}^{(n)}\sim O(z^{-n+\ell}). \eqno\eq
$$
Then the conversion of $\xi_{\ell m}$ to $X^{in}_{\ell m\omega}$
by Eq.\xixi\ yields the result Eq.$\xasy$.
Since $\epsilon=O(v^3)$ and $z=O(v)$,
in order to calculate the energy and angular momentum fluxes
up to $O(v^5)$ beyond Newtonian, we conclude that
the necessary power series formulas for $X^{(n)}_{\ell m\omega}$ around
$z=0$ are those of $n\leq2$ for $\ell=2$, $n\leq1$ for $\ell=3$
and $n=0$ for $\ell=4$.

\Appendix{D}

In this Appendix we describe the expansion of the spheroidal harmonics
${}_{-2}S_{\ell m}^{a\omega}$ and their eigenvalues $\lambda$
in powers of $a\omega$.
Since we are interested in the energy and angular momentum fluxes
to $O(v^5)$ in this paper,
we need the expansions of ${}_{-2}S_{\ell m}^{a\omega}$
to $O(a\omega)$ for both $\ell=2$ and $3$, but
those of $\lambda$  to $O((a\omega)^2)$ for $\ell=2$ and
to $O(a\omega)$ for $l=3$, while we only need
the lowest order formulas for $\ell=4$.

The spheroidal harmonics of spin weight $s=-2$ obey the equation,
$$
\eqalign{
 \Bigl[{1 \over \sin\theta}{d \over d\theta}
 \Bigl\{\sin\theta {d \over d\theta} \Bigr\}
-&a^2\omega^2\sin^2\theta
 -{(m-2\cos\theta)^2 \over \sin^2\theta}
\cr
&+4a\omega\cos\theta-2+2ma\omega+\lambda\Bigr]
{}_{-2}S_{\ell m}^{a\omega}=0.\cr
}\eqn\esph
$$
We assume they are normalized according to Eq.\Snorm.
We expand ${}_{-2}S_{\ell m}^{a\omega}$ and $\lambda$ as
$$
\eqalign{
{}_{-2}S_{\ell m}^{a\omega}
&={}_{-2}P_{\ell m}+a\omega S_{\ell m}^{(1)}
+(a\omega)^2 S_{\ell m}^{(2)}+O((a\omega)^3),
\cr
\lambda&=\lambda_0+a\omega\lambda_1+a^2\omega^2 \lambda_2
+O((a\omega)^3),
\cr}
\eqn\seq
$$
where ${}_{-2}P_{\ell m}$ are the spherical harmonics of spin weight
$s=-2$, $\lambda_0=(\ell-1)(\ell+2)$ and
$\lambda_1=-2m(\ell^2+\ell+4)/(\ell^2+\ell)$\rlap.\refmark\press\

Inserting Eq.\seq\ into Eq.\esph\ and collecting the terms of equal
orders in $a\omega$ and $(a\omega)^2$, we obtain
$$
\eqalignno{
&{\cal L}_0 S_{\ell m}^{(1)}+\lambda_0 S_{\ell m}^{(1)}
=-(4\cos\theta+2m+\lambda_1)\,_{-2}P_{\ell m}\,,
&\eqnalign\ieq\cr
&{\cal L}_0 S_{\ell m}^{(2)}+\lambda_0 S^{(2)}_{\ell m}
=-(4\cos\theta+2m+\lambda_1)S_{\ell m}^{(1)}
-(\lambda_2-\sin^2\theta)~_{-2}P_{\ell m}\,,
&\eqnalign\jeq\cr}
$$
where ${\cal L}_0$ is the operator for the
spin-weighted spherical harmonics,
$$
\eqalign{
{\cal L}_0[{}_{-2}P_{\ell m}]
& \equiv\left[ {1 \over \sin\theta}{d \over d\theta}
\Bigl\{ \sin\theta {d \over d\theta} \Bigr\}
-{(m-2\cos\theta)^2 \over \sin^2\theta}-2\right]
{}_{-2}P_{\ell m}\cr
&=-\lambda_0~{}_{-2}P_{\ell m}\,.\cr}
\eqn\eyph
$$

First we solve Eq.\ieq\ for $S_{\ell m}^{(1)}$. Setting
$$
S_{\ell m}^{(1)}=\sum_{\ell'} c_{\ell m}^{\ell'}~{}_{-2}P_{\ell'm}\,,
\eqn\llll
$$
we insert it into Eq.\ieq, multiply it by ${}_{-2}P_{\ell'm}$
and integrate it over $\theta$.
Then noting the normalization of the spheroidal harmonics, we have
$$
c_{\ell m}^{\ell'}=
\cases{\displaystyle
{4 \over (\ell'-1)(\ell'+2)-(\ell-1)(\ell+2)}\int
\,{}_{-2}P_{\ell'm}\cos\theta\, {}_{-2}P_{\ell m} d\cos\theta\,,
\quad&$\ell' \not=\ell$,
\crr
0,&$\ell'=\ell$.
\cr}
\eqno\eq
$$
Hence $c_{\ell m}^{\ell'}$ is non-zero only for $\ell'=\ell \pm 1$, and
we obtain
$$\eqalign{
&c_{\ell m}^{\ell+1}={2 \over (\ell+1)^2}\Bigl\lbrack
{(\ell+3)(\ell-1)(\ell+m+1)(\ell-m+1) \over (2\ell+1)(2\ell+3)}
 \Bigr\rbrack^{1/2},\cr
&c_{\ell m}^{\ell-1}=-{2 \over \ell^2}\Bigl\lbrack
{(\ell+2)(\ell-2)(\ell+m)(\ell-m) \over (2\ell+1)(2\ell-1)}
\Bigr\rbrack^{1/2}.\cr}
\eqno\eq
$$

Next we consider $\lambda_2$.
We set
 $S_{\ell m}^{(2)}=\sum_{\ell'} d_{\ell m}^{\ell'}~{}_{-2}P_{\ell'm}$
and insert it into Eq.\jeq. This time
we multiply it by ${}_{-2}P_{\ell m}$ and integrate it over $\theta$,
noting that $d^\ell_{\ell m}=0$.
The result is
$$\eqalign{
\lambda_2
&=-4\int{}_{-2}P_{\ell m}\cos\theta\, S_{\ell m}^{(1)} d\cos\theta
+\int{}_{-2}P_{\ell m}\sin^2\theta\,{}_{-2}P_{\ell m} d\cos\theta\cr
&=-2(\ell+1)(c_{\ell m}^{\ell+1})^2+2\ell(c_{\ell m}^{\ell-1})^2
+1-\int{}_{-2}P_{\ell m}\cos^2\theta\,{}_{-2}P_{\ell m} d\cos\theta.\cr}
\eqno\eq
$$
The last integral becomes
$$
\int{}_{-2}P_{\ell m}\cos^2\theta\,{}_{-2} P_{\ell m} d\cos\theta=
{1 \over 3}+{2 \over 3}{(\ell+4)(\ell-3)(\ell^2+\ell-3m^2) \over
\ell(\ell+1)(2\ell+3)(2\ell-1)}\,.
\eqno\eq
$$
We need $\lambda_2$ only for $\ell=2$. In this case, the final answer
becomes
$$
\lambda_2(\ell=2)={90-10m^2 \over 189}\,.\eqno\eq
$$

\refout
\end